\documentstyle[11pt,aaspp4]{article}
\def\lineunits{ergs\ s$^{-1}$\,cm$^{-2}$}
\def\contunits{ergs\ s$^{-1}$\,cm$^{-2}$\,\AA$^{-1}$}
\def\Fvar{\ifmmode F_{\rm var} \else $F_{\rm var}$\fi}
\def\Rmax{\ifmmode R_{\rm max} \else $R_{\rm max}$\fi}
\def\tcent{\ifmmode \tau_{\rm cent} \else $\tau_{\rm cent}$\fi}
\def\IUE{{\it IUE}}

\def\arcsec{$''$}

\def\arcsecpoint{$''\!.$}

\def\kms{\ifmmode {\rm km\ s}^{-1} \else km s$^{-1}$\fi}
\def\Msun{\ifmmode M_{\odot} \else $M_{\odot}$\fi}
\def\Lsun{\ifmmode L_{\odot} \else $L_{\odot}$\fi}
\def\qo{\ifmmode q_{\rm o} \else $q_{\rm o}$\fi}
\def\Ho{\ifmmode H_{\rm o} \else $H_{\rm o}$\fi}
\def\ho{\ifmmode h_{\rm o} \else $h_{\rm o}$\fi}

\def\vFWHM{\ifmmode v_{\mbox{\tiny FWHM}} \else
            $v_{\mbox{\tiny FWHM}}$\fi}
\def\CCF{\ifmmode F_{\it CCF} \else $F_{\it CCF}$\fi}
\def\ACF{\ifmmode F_{\it ACF} \else $F_{\it ACF}$\fi}
\def\Halpha{\ifmmode {\rm H}\alpha \else H$\alpha$\fi}
\def\Hbeta{\ifmmode {\rm H}\beta \else H$\beta$\fi}
\def\Hgamma{\ifmmode {\rm H}\gamma \else H$\gamma$\fi}
\def\Hdelta{\ifmmode {\rm H}\delta \else H$\delta$\fi}
\def\Lya{\ifmmode {\rm Ly}\alpha \else Ly$\alpha$\fi}
\def\Lyb{\ifmmode {\rm Ly}\beta \else Ly$\beta$\fi}

\def\ciii{\ifmmode {\rm C}\,{\sc iii} \else C\,{\sc iii}\fi}
\def\civ{\ifmmode {\rm C}\,{\sc iv} \else C\,{\sc iv}\fi}

\def\oiii{O\,{\sc iii}}
\def\o5007{[O\,{\sc iii}]\,$\lambda5007$}

\def\feii{Fe\,{\sc ii}}

\lefthead{Peterson et al.}
\righthead{Long-Term Monitoring of NGC 5548}

\begin{document}
\title{Steps Toward Determination of the Size and Structure
of the Broad-Line Region in Active Galactic Nuclei.\ XV.\
Long-Term Optical Monitoring of NGC~5548}

\author{
B.M.~Peterson,\altaffilmark{1}
A.J.~Barth,\altaffilmark{2}
P.~Berlind,\altaffilmark{3}
R.~Bertram,\altaffilmark{1}$^{,}$\altaffilmark{4}
K.~Bischoff,\altaffilmark{5}
N.G.~Bochkarev,\altaffilmark{6}
A.N.~Burenkov,\altaffilmark{7}
F.-Z.~Cheng,\altaffilmark{8}
M.~Dietrich,\altaffilmark{9}
A.V.~Filippenko,\altaffilmark{2}
E.~Giannuzzo,\altaffilmark{10}
L.C.~Ho,\altaffilmark{3}
J.P.~Huchra,\altaffilmark{3}
J.~Hunley,\altaffilmark{1}
S.~Kaspi,\altaffilmark{11}
W.~Kollatschny,\altaffilmark{5}
D.C.~Leonard,\altaffilmark{2}
Yu.F.~Malkov,\altaffilmark{12}
T.~Matheson,\altaffilmark{2}
M.~Mignoli,\altaffilmark{13}
B.~Nelson,\altaffilmark{14}
P.~Papaderos,\altaffilmark{5}
J.~Peters,\altaffilmark{3}$^{,}$\altaffilmark{15}
R.W.~Pogge,\altaffilmark{1}
V.I.~Pronik,\altaffilmark{12}$^{,}$\altaffilmark{16}
S.G.~Sergeev,\altaffilmark{12}$^{,}$\altaffilmark{16}
E.A.~Sergeeva,\altaffilmark{12}$^{,}$\altaffilmark{16}
A.I.~Shapovalova,\altaffilmark{7}
G.M.~Stirpe,\altaffilmark{13}
S.~Tokarz,\altaffilmark{3}
R.M.~Wagner,\altaffilmark{1}$^{,}$\altaffilmark{4}
I.~Wanders,\altaffilmark{1}$^{,}$\altaffilmark{17}
J.-Y.~Wei,\altaffilmark{18}
B.J.~Wilkes,\altaffilmark{3}
H.~Wu,\altaffilmark{18}
S.-J.~Xue,\altaffilmark{8}
and Z.-L.~Zou\,\altaffilmark{18}
}
\altaffiltext{1}
        {Department of Astronomy, The Ohio State University,
		174 West 18th Avenue, Columbus, OH  43210--1106.
	\\ Email: peterson, pogge@astronomy.ohio-state.edu,
	jhunley@elmers.com}
\altaffiltext{2}
	{Department of Astronomy, University of California at Berkeley,
	Berkeley, CA 94720--3411.
	\\ Email: abarth, afilippenko, dleonard, tmatheson@astro.berkeley.edu}
\altaffiltext{3}
	{Harvard-Smithsonian Center for Astrophysics,
	60 Garden Street, Cambridge, MA 02138.
	\\ Email: pberlind, lho, huchra, belinda@cfa.harvard.edu}
\altaffiltext{4}
	{Mailing address: Lowell Observatory, 1400 West Mars Hill Road, 
	Flagstaff, AZ 86001.
	\\ Email: rayb@lowell.edu, rmw@altair.lowell.edu}
\altaffiltext{5} 
	{Universit\"{a}ts-Sternwarte G\"{o}ttingen, Geismarlandstr.\ 11,
	D--37083 G\"{o}ttingen, Germany.
	\\ Email: bischoff, papade@uni-sw.gwdg.de, wkollat@gwdg.de}
\altaffiltext{6}
	{Sternberg Astronomical Institute, University of Moscow,
	Universitetskij Prosp.\ 13, Moscow 119899, Russia.
	\\ Email: boch@astronomy.msk.su}
\altaffiltext{7}
	{Special Astrophysical Observatory, Russian Academy of Sciences,
	Nizhnij Arkhyz, Karachai-Cherkess Republic, 357147, Russia.
	\\ Email: ashap@sao.ru}
\altaffiltext{8}
	{Center for Astrophysics, University of Science and Technology
	of China, Hefei, Anhui, People's Republic of China.
	\\ Email: fzhen@ustc.edu.cn}
\altaffiltext{9}
	{Landessternwarte Heidelberg, K{o}nigstuhl, 
	D--69117 Heidelberg, Germany.
	\\ Email: mdietric@mail.lsw.uni-heidelberg.de}
\altaffiltext{10}
	{Department of Physical Sciences, University of Hertfordshire,
	College Lane, Hatfield, HERTS AL10 9AB, United Kingdom.
	\\ Email: ester@star.herts.ac.uk}
\altaffiltext{11}
	{School of Physics and Astronomy and the Wise Observatory,
	The Raymond and Beverly Sackler Faculty of Exact Sciences,
	Tel-Aviv University, Tel-Aviv 69978, Israel.	
	\\ Email: shai@wise.tau.ac.il}
\altaffiltext{12}
	{Crimean Astrophysical Observatory, P/O Nauchny, 334413 Crimea,
	Ukraine. 
	\\ Email: vpronik, sergeev, selena@crao.crimea.ua}
\altaffiltext{13}
	{Osservatorio Astronomico di Bologna,
	Via Zamboni 33, I--40126 Bologna, Italy.
	\\ Email: mignoli, stirpe@astbo3.bo.astro.it}
\altaffiltext{14}
	{Department of Physics and Astronomy, University of California
	at Los Angeles, Math-Science Building, Los Angeles, CA 90024.
	\\ Email: nelson@astro.ucla.edu}
\altaffiltext{15}
	{Deceased.}
\altaffiltext{16}
	{Also: Isaac Newton Institute of Chile, Crimean Branch.}
\altaffiltext{17}
	{Mailing address: Mercatorpad 4, bus 401, 3000 Leuven, Belgium
	\\ Email: xenu@xs4all.nl}
\altaffiltext{18}
	{Beijing Astronomical Observatory, Beijing 100080,
	People's Republic of China.
	\\ Email: wu@bac.pku.edu.cn, zouzl@sun.ihep.ac.cn}

\clearpage

\begin{abstract}
We present the results of three years of ground-based observations of
the Seyfert 1 galaxy NGC 5548, which combined with previously
reported data, yield optical continuum and broad-line \Hbeta\
light curves for a total of eight years. The light curves
consist of over 800 points, with a typical spacing of a few days
between observations. During this eight-year period, the 
nuclear continuum has varied by more than a factor of
seven, and the \Hbeta\ emission line has varied by
a factor of nearly six.
The \Hbeta\ emission line responds 
to continuum variations with a time delay or lag of
$\sim10$--20\,days, the precise value varying somewhat from
year to year. We find some indications that the lag
varies with continuum flux in the sense that the lag
is larger when the source is brighter.
\end{abstract}

 
\setcounter{footnote}{0}

\section{Introduction}

The observed time delays between continuum and emission-line
flux variations can be used to
determine the size and structure of the broad-line region (BLR) in 
active galactic nuclei (AGNs), through the process known
as ``reverberation mapping'' (Blandford \& McKee 1982).
It is assumed that the continuum light curve $C(t)$
and emission-line light curve $L(t)$ are related as
\begin{equation}
\label{eq:TF}
L(t) = \int \Psi(\tau)\ C(t-\tau)\ d\tau,
\end{equation}
where $\Psi(t)$ is the ``transfer function,'' which depends on
the geometry and reprocessing physics of the BLR. The aim
of reverberation mapping is to use the observed light curves
$C(t)$ and $L(t)$ to solve for $\Psi(\tau)$ and thus infer
the geometry  of the BLR. On account of the amount and quality
of data required to determine $\Psi(\tau)$ uniquely,
reverberation-mapping studies are still in a state of relative
infancy, and it is far more common to obtain only a cross-correlation
function (CCF) $\CCF(\tau)$ between the continuum and emission-line light
curves. By convolving eq.\ (\ref{eq:TF}) with $C(t)$, it is simple
to show that 
\begin{equation}
\CCF(\tau) = \int \Psi(\tau') \ACF(\tau-\tau')\ d\tau',
\end{equation}
where $\ACF(\tau)$ is the continuum autocorrelation function
(Penston 1991; Peterson 1993). It can also be shown that the
centroid of the CCF is a measure of the emissivity-weighted mean
radius of the variable part of the BLR (Koratkar \& Gaskell 1991). 
Therefore even relatively unsophisticated time-series analysis can give
us an appropriate scale length for the BLR by simply 
measuring the time-delayed response of an emission line to 
continuum variations and interpreting this time delay as due
to light travel-time effects within the BLR.

Over the last decade, high-intensity spectroscopic monitoring programs
have led to determination of characteristic emission-line
response times for about two dozen mostly low-luminosity AGNs
(see Netzer \& Peterson 1997 for a recent review).
The most extensive of these various efforts is
a continuing program of optical ground-based monitoring of the
bright Seyfert galaxy NGC 5548 that has been carried out by
the International AGN Watch consortium (Alloin et al.\ 1994).
This program was initiated in late 1989 (Peterson et al.\ 1991)
in support of a massive {\it International Ultraviolet Explorer (IUE)}\  
monitoring program
(Clavel et al.\ 1991). The ground-based program has continued
uninterrupted since then, and also supported a
second large multiwavelength
project involving both {\it Hubble Space Telescope} and \IUE\
(Korista et al.\ 1995), as well as the
{\it Extreme Ultraviolet Explorer}\ (Marshall et al.\ 1997).
Five years of ground-based data on optical continuum
(at 5100\,\AA) and broad \Hbeta\ emission-line variations 
in NGC 5548 have already
been published (Peterson et al.\ 1991, 1992, 1994;
Korista et al.\ 1995).
In this contribution, we present
three additional years of ground-based optical data on
this source.

The principal aim of continuing this program is to search
for changes in the BLR on a dynamical time scale that might
be manifest in changes in the \Hbeta\ response time or
in the emission-line profiles. This program also provides
well-sampled continuum observations over an extended
period, which may ultimately provide important clues about
the origin of the continuum and its variability.

As is customary for the papers in this series, we will
focus primarily on the data and the salient results of
cross-correlation analysis of the continuum and emission-line
light curves. Further in-depth analysis will be left to other papers.
In \S\,2 we describe the observations, data reduction, and
intercalibration procedures that we have used to
construct a homogeneous data base of optical continuum
and \Hbeta\ emission-line fluxes. In \S\,3, we describe
the time-series analysis that we have undertaken
to determine the time scale for response of \Hbeta\
to continuum variations. Our results are summarized in \S\,4.

\section{Observations and Data Analysis}

\subsection{Spectroscopic Observations}

Spectroscopic observations were obtained 
between UT 1993 November 17 (Julian Date = JD2449309) and
1996 October 16 (JD2450373), covering three 
separate observing seasons on NGC 5548. 
Throughout this
paper we will refer to the interval
1993 November to 1994 October (JD2449309 to JD2449636) as Year 6,
1994 November to 1995 October (JD2449679 to JD2450008) as Year 7, and
1995 November to 1996 October (JD2450044 to JD2450373) as Year 8.
During this three-year period, a total of 309 spectra covering
at least the \Hbeta\ region of the spectrum 
were obtained with 
CCD spectrographs on a wide variety of telescopes, as summarized 
in Table 1. Column (1) gives a code for each data set; these
are the same codes that have been used throughout the 
NGC 5548 monitoring project.
A more complete log of observations, as has been
published in our previous papers on NGC 5548, can be obtained
at the International AGN Watch 
site on the World-Wide Web\footnote{The light curves 
and complete logs of observation are
available in tabular form at URL 
{\sf http://www.astronomy.ohio-state.edu/$\sim$agnwatch/}.
All publicly available International AGN Watch data can be accessed
at this site, which also includes complete references to
published AGN Watch papers.}.

The spectroscopic images were processed by individual observers
in standard
fashion for CCD frames, including bias subtraction, 
flat-field correction, wavelength calibration,
and flux calibration based on standard-star observations.
However, the usual technique of flux calibration by
comparison with standard stars is far too poor for
AGN variability studies, since even under nominally
photometric conditions, AGN spectrophotometry is rarely more
accurate than $\sim$10\%. We therefore rely on standard-star
calibration {\em only}\ for relative calibration, i.e., relative
flux as a function of wavelength. Absolute calibration is
based on the prominent narrow \o5007\ emission line. The
narrow emission lines arise in a lower-density region that is quite compact,
but still significantly larger than the BLR
(Kraemer et al.\ 1998; Sergeev et al.\ 1997). The larger spatial extent and
long recombination time mean that \o5007\ flux variations on the
time scales discussed here will be undetectable. We therefore
assume that the \o5007\ line remains constant in flux,
and can be used as an internal flux calibrator. All spectra
are scaled to a constant flux of
$F(\mbox{\o5007}) = 5.58 \times 10^{-13}$\,\lineunits.
This absolute flux was determined from spectra made under photometric
conditions during the first year of this program 
(Peterson et al.\ 1991). We have also measured the
\o5007\ flux for the new spectra that were obtained 
through large spectrograph entrance apertures under photometric
conditions, and these results are given in Table 2.
This is a conservative list, in that have we included only data
where the observer recorded that the observing conditions appeared
to be of photometric quality.
These measurements suggest that the calibration that we have used
for all eight years might be too high by no more than a few percent,
but there is no real evidence for time variability
of the \o5007\ flux. We note, however, that some evidence for
low-amplitude variability of \o5007\ on time scales of
single observing seasons ($\sim 200$\,days) has been claimed
(Sergeev et al.\ 1997),
but this has little if any effect on the results reported here.

Once a standard \o5007\ flux has been adopted, all of the
spectra are adjusted in flux to have this value. We have
done this by employing the spectral scaling software described
by van Groningen \& Wanders (1992). We adopt a high-quality
spectrum that has been scaled by a multiplicative constant to
the adopted \o5007\ flux as a ``reference spectrum,'' and all
other spectra are scaled relative to it in a least-squares
fashion that minimizes the [\oiii] residuals in the difference
spectrum produced by subtracting the reference spectrum from
each program spectrum. This  
program also corrects for small zero-point
wavelength-calibration errors between the
individual spectra, and takes resolution differences into account. 

At this point, measurements of each of the spectra are made.
The continuum flux at $\sim5100$\,\AA\ (in the rest frame of
NGC 5548, $z = 0.0167$) is determined by averaging the flux in 
the 5185--5195\,\AA\ bandpass (in the observed frame).
The \Hbeta\ emission-line flux is measured by assuming a straight
underlying continuum between $\sim4790$\,\AA\ and $\sim5170$\,\AA,
and integrating the flux above this continuum between 4795\,\AA\ and
5018\,\AA\ (all wavelengths in the observed frame). The long-wavelength
cutoff of this integration band misses some of the \Hbeta\ flux
underneath [\oiii]\,$\lambda4959$, but avoids the need to
estimate the \feii\ contribution to this feature and still gives
a good representation of the \Hbeta\ variability. We also note
that no attempt has been made to correct for contamination of
the line measurement by the {\em narrow-line}\ component of 
\Hbeta, which is of course expected to be constant. Recent
estimates of the strength of narrow-line \Hbeta\ range
between $0.12F(\mbox{\o5007})$ (Kraemer et al.\ 1998) and
$0.151F(\mbox{\o5007})$ (Wanders \& Peterson 1996), in any case
typically only $\sim10$\% of the measured \Hbeta\ flux.

Even after scaling all of the spectra to a common
value of the \o5007\ flux, there are systematic differences
between the light curves produced from data
obtained at different telescopes. As in our previous papers,
we correct for the small offsets between the light curves from
different sources in a simple, but effective, fashion. We
attribute these small relative offsets
to aperture effects (Peterson et al.\ 1995), although the
procedure we use also corrects for other unidentified systematic
differences between data sets.  We
define a point-source correction factor $\varphi$ by the equation
\begin{equation}
\label{eq:defphi}
F(\Hbeta)_{\rm true} = \varphi F(\Hbeta)_{\rm observed}.
\end{equation}
This factor accounts for the fact that different apertures
result in different amounts of light loss for the 
point-spread function (which describes the surface-brightness
distribution of both the broad lines and the AGN continuum
source) and the partially extended narrow-line region.
We note, of course, that this correction factor is in
fact a function of seeing (Peterson et al.\ 1995). 
We do not attempt to
correct for seeing effects, and this is probably our largest single
source of uncertainty. 

After correcting for aperture effects on the point-spread function
to narrow-line ratio, another correction needs to be applied to
adjust for the different amounts of starlight admitted by
different apertures. An extended source correction $G$ 
is thus defined as
\begin{equation}
\label{eq:defG}
F_{\lambda}(5100\,{\textstyle {\rm \AA}})_{\rm true} = \varphi 
F_{\lambda}(5100\,{\textstyle {\rm \AA}})_{\rm observed} - G.
\end{equation}

Intercalibration of the various data
sets is then accomplished  by
comparing pairs of nearly simultaneous observations from different
data sets to determine 
for each data set the values of the constants $\varphi$ and $G$
which are needed to adjust the emission-line and continuum
fluxes to a common scale. Furthermore, the formal uncertainties
in $\varphi$ and $G$ reflect the uncertainties in the 
individual data sets, so we can determine the nominal
uncertainties for each data set if we assume that the errors add
in quadrature. In practice, the interval which we define
as ``nearly simultaneous'' is typically one or two days,
which means that any real variability that occurs on time scales
this short  tends to be somewhat suppressed by the
process that allows us to merge the different data sets.

As in our previous work, 
the data are adjusted  relative to data set ``A'' 
because these data are fairly extensive, 
overlap well with most of the other data sets, and were obtained through
a reasonably large aperture (5\arcsecpoint0 $\times$ 7\arcsecpoint5).
Fractional uncertainties 
of $\sigma_{cont}/F_{\lambda}(5100$\,\AA) $\approx 0.020$ and
$\sigma_{line}/F(\Hbeta) \approx 0.020$ for the continuum
and \Hbeta\ line, respectively, are adopted for the similar, 
large-aperture, high-quality data sets ``A'' and ``H,''
based on the differences between closely spaced observations
within these sets (see also Peterson et al.\ 1998a). 
For the other data sets, it was possible to
estimate the mean uncertainties in the measurements by
comparing them to measurements from other sets for
which the uncertainties are known and by assuming that
the uncertainties for each set add in quadrature.

The intercalibration constants we use for each data set are
given in Table 3, and these constants are used with 
equations (\ref{eq:defphi}) and (\ref{eq:defG}) to adjust
the spectral measurements. We note that it was not possible
to find a single set of correction factors for 
the data in set ``F,'' probably on account of modifications to the
instrument made during this period 
(Fabricant et al.\ 1998). We have thus determined
separate intercalibration constants for each of the three years.

The final continuum [$F_{\lambda}$(5100\,\AA)] and \Hbeta\ 
emission-line fluxes are given in Table 4. Simultaneous
(to within 0.1 day) measurements were averaged, weighted by
the reciprocal of their variances.
We can perform a final check of our
uncertainty estimates by examining the ratios of all pairs of
observations in Table 4 which are 
separated by 1 day or less. There are 160 independent
pairs of measurements within 1 day.
The dispersion about the mean (unity),
divided by $\sqrt{2}$, provides an estimate of the typical uncertainty
in a single measurement.
For the continuum, we find that the mean fractional
error in a given measurement is 0.032. The 
average fractional uncertainty, from the quoted estimates 
for these same 160 pairs of measurements, is 0.035.
Similarly, for the \Hbeta\ light curve, 
we find that the mean fractional
error in a given measurement is 0.031. The 
average fractional uncertainty  from the quoted estimates 
for these same 160 pairs of measurements is 0.035.
These numbers imply that our error estimates are probably quite good.
The mean fractional errors in this data set are somewhat
superior to those quoted in our earlier papers, and this is probably
due to the fact that, in contrast to the past,
{\em all} of the data reported here were obtained
with CCD detectors.

\subsection{Photometric Observations}

In addition to the spectroscopic observations, a limited
amount of CCD photometric data also have been obtained.
The $V$-band measurements obtained from these observations
are given in Table 5, measured by using 
simulated aperture photometry with a circular 
aperture of diameter 10\arcsec. In each case, the $V$ magnitudes 
given in column (2) are slightly offset from the continuum
measured from the spectroscopic data. The data have been
adjusted and converted to fluxes $F_{\lambda}$(5100\,\AA)
(as given in column 4) using the procedure described in
detail by Korista et al.\ (1995); the systematic corrections
employed (eq.\ 4 of Korista et al.) are $\Delta m = -0.254\pm0.048$\,mag
for data set ``A'' and $\Delta m = -0.085\pm0.050$\,mag
for sets ``B'' and ``C.''

The final light curves for the optical continuum and \Hbeta\ fluxes
during Years 6 through 8 are shown in Fig.\ 1. The continuum points
are from column (2) of Table 3 and column (4) of Table 5,
and the \Hbeta\ fluxes are those given in column (3) of Table 3.
All measurements are in the observer's
frame, and are uncorrected for Galactic extinction.

\section{Variability Analysis}

The data shown in Fig.\ 1 can be combined with our previously
published light curves (Peterson et al.\ 1991, 1992, 1994;
Korista et al.\ 1995), yielding combined homogeneous light
curves that cover a span of 2865 days. The combined data are shown in Fig.\ 2.
These light curves are comprised of 857 continuum measurements
and 820 line measurements. The number of spectra in this data base
exceeds by more than a factor of
four the number of spectra in any other homogeneous
time series on UV or optical variations
in Seyfert galaxies (cf.\ Wanders et al.\ 1997; Peterson et al.\ 1998a),
and previous less-well sampled and less-homogenous data extend
the data base on this object to 20 years (Sergeev et al.\ 1997).
In this section, we will summarize the basic characteristics of
the eight-year homogeneous data base,  both in its entirety
and on a year-by-year basis.

\subsection{Characteristics of the Data Base}

Tables 6 and 7 provide a summary, for the continuum and \Hbeta\ emission
line respectively, of the basic characteristics of
the light curves of NGC 5548, broken down into subsets of time
as given in column (1). The number of observations in the
subset is given in column (2), and columns (3) and (4) give
the average and median intervals, 
respectively, between successive observations.
Note that for the entire data set, the relatively large ratio of
the average to the median is attributable to the one- to two-month gaps
between observing seasons. The mean and 
root-mean-square (rms) fluxes, $F_{\lambda}$(5100\,\AA)
in Table 6 and $F$(\Hbeta) in Table 7, are given in column (5).
Two standard measures of variability, \Fvar\ and \Rmax, are given
in columns (6) and (7), respectively. 
The parameter \Fvar\ is the rms fractional variability,
corrected for measurement error, as defined by 
Rodr\'{\i}guez-Pascual et al.\ (1997), and \Rmax\ is simply
the ratio of maximum to minimum flux. Both of these parameters
are affected by contamination of the measured quantities by
constant-flux components, the underlying host galaxy in
the case of the continuum, and the narrow \Hbeta\ emission
component in the case of the line. The continuum flux measurements
are all adjusted to a standard entrance aperture of
5\arcsecpoint0$\times$7\arcsecpoint5, through which the starlight
component can be estimated to contribute $V_{\rm galaxy} = 14.99$\,mag, or 
$F_{\lambda}$(5100\,\AA)$_{\rm galaxy} = 3.4 \times 10^{-15}$\,\contunits\
(Romanishin et al.\ 1995).
Subtracting this value from all the continuum points in Fig.\ 2
increases \Fvar\ from 0.195 to 0.307 and \Rmax\ from 
$2.98\pm0.16$ to $7.26\pm0.98$, i.e., the
AGN nuclear continuum has varied by more than a factor of 7 in eight years.
Similarly, we can estimate that the narrow \Hbeta\ contribution to
the line measurements is about $8.43 \times 10^{-14}$\,\lineunits\
(Wanders \& Peterson 1996).
Subtracting this from the line measurements yields
$\Fvar = 0.224$ and $\Rmax = 5.67\pm0.56$ for the broad \Hbeta\ component.

During Year 7, NGC 5548 achieved the highest luminosity state we have
observed. The faintest state observed was during Year 4. In Fig.\ 3,
we show set ``H''
spectra obtained near the times of extreme states. Note in particular
the dramatic difference in the continuum levels shortward of about
4200\,\AA, the onset of the Balmer continuum and \feii\ emission-line
blend referred to as the ``small blue bump'' (see Maoz et al.\ 1993
for a discussion of the variability of the small blue bump in
NGC 5548). The low-state spectrum in Fig.\ 3 shows that the
narrow-line components of the Balmer lines are quite strong;
the high-state spectrum is in fact strongly double-peaked, but the
central depression is largely filled in by the narrow component. This
is why the double-peaked structure of the broad lines is much more
prominent in the difference spectrum than in the high-state spectrum.

\subsection{Time-Series Analysis}

Inspection of Fig.\ 1 shows that there is a clear time delay between
continuum variations and the \Hbeta\ response. This time delay
can be quantified by cross-correlation of the continuum and emission-line
light curves. We perform the cross-correlation analysis in two
ways, using the interpolation cross-correlation function (ICCF) method
of Gaskell \& Sparke (1986) and Gaskell \& Peterson (1987) and the
discrete correlation function (DCF) method of Edelson \& Krolik (1988).
In both cases, we use the implementation described by White \& Peterson
(1994). 

The results of the cross-correlation analysis are shown in
Fig.\ 4 and summarized in Table 8. In Table 8, the subset used in
cross-correlation is given in column (1). Column (2) gives the
centroid \tcent\ of the ICCF. The highest point
in the ICCF occurs at a time delay
$\tau_{\rm peak}$ (column 3) and has value $r_{\rm max}$
(column 4). The full-width at half-maximum (FWHM) of the ICCF 
is given in column (5). The centroid \tcent\
is computed using all points near $\tau_{\rm peak}$ with
values greater than 0.8$r_{\rm max}$. 
The uncertainties quoted for \tcent\ and 
$\tau_{\rm peak}$ have been computed using the model-independent
FR/RSS Monte-Carlo method described by Peterson et al.\ (1998b).

\section{Discussion}

It is apparent from inspection of Table 8 
that significant changes in the \Hbeta\ lag have occurred
during this eight-year monitoring program. The cross-correlation
functions shown in Fig.\ 4 show pronounced changes from year to year.
The centroid \tcent\ is of particular
interest as it is a measure of the responsivity-weighted mean radius
of the BLR (Koratkar \& Gaskell 1991; Penston 1991).
Possible changes in the \Hbeta\ lag have been noticed previously, and
might plausibly arise in a number of ways:
\begin{enumerate}
\item A physically thick BLR has a broad transfer function,
and the centroid convolution of $\Psi(\tau)$ and $\ACF(\tau)$
can become quite sensitive to the general character of the
continuum variations as embodied in the ACF (Netzer \& Maoz 1990).
\item Real dynamical changes in the BLR might occur on time
scales of years (Wanders 1995). This has also been suggested
to be the case in NGC 4151 (e.g., Ulrich et al.\ 1991).
\item The lag might vary with the ionizing flux incident upon the
BLR clouds. For BLR clouds with otherwise constant parameters,
the radius at which emissivity in a particular line is optimized
will be dependent on the luminosity of the source
(e.g., Baldwin et al.\ 1995).
\end{enumerate}

In Fig.\ 5, we show in the upper panel \tcent\ as a function
of time for the eight years of International AGN Watch data
reported here, plus the results of the 1988 monitoring program
carried out at Wise Observatory\footnote{We have adjusted the
Wise Observatory data  to form as nearly as possible
a quasi-homogeneous
set with the International AGN Watch data. Through simulated aperture
photometry based on the nucleus-free image described by Romanishin
et al.\ (1995), we estimate that the starlight contribution through
the Wise Observatory entrance aperture is 
$1.166\times10^{-14}$\,\contunits. The 5375--5575\,\AA\ fluxes of
Netzer et al.\ have been adjusted to the nominal galaxy contribution
in the AGN Watch data, $3.4\times10^{-15}$\,\contunits. The 
cross-correlation methodology used in this paper gives
$\tcent = 8.44^{+4.01}_{-3.87}$ for the Wise Observatory campaign.}
(Netzer et al.\ 1990).
This bears some resemblance to the light curve
averaged over individual observing seasons, as shown in the
lower panel of the figure, suggesting that the lag may be
a luminosity-dependent quantity. In Fig.\ 6, we plot
directly \tcent\ as a function $\langle F_{\lambda}({\rm 5100\,\AA})\rangle$,
the mean continuum flux. Again, this suggests a relationship
between these quantities. The  slope of this
relationship is formally non-zero at $2.5\sigma$
significance.

\section{Summary}

We have presented new optical spectroscopic and photometric
observations of the continuum and \Hbeta\ emission-line
variations in the Seyfert 1 galaxy NGC 5548, thus extending
our high-intensity coverage of this object to eight years.
The total continuum light curve consists of 857 epochs and
the \Hbeta\ light curve consists of 825 epochs. We find as
before that the variations of the \Hbeta\ emission line
follow those of the continuum by $\sim10$--20\,days, with
the precise value of the measured lag changing somewhat from 
year to year. There are indications that the lag varies with
the continuum flux in the sense that the lag is larger when
the continuum is brighter.
\bigskip

We are grateful to the Directors and Telescope Allocation
Committees of our various observatories for their support of
this project. Individual investigators have benefited from
the support from a number of grants, including the following:
National Science Foundation grants AST--9420080 (Ohio State University)
and AST--9417213 (University of California at Berkeley);
Sonderforschungsbereich grant 328 (Landessternwarte Heidelberg),
DFG grant Ko 857/13-2 (Universit\"{a}ts-Sternwarte G\"{o}ttingen),
the Russian Basic Research Foundation grants 94--02---4885 and 97--02--17625;
and the PanDen Project of the Chinese National Committee of Sciences and
the Chinese National Science Foundation.



\clearpage

\begin{figure}
\caption{The optical continuum (upper panel) and \Hbeta\
emission-line light curves for NGC 5548 between
1993 November and 1996 October. These are based on 309
optical spectra and 14 photometric observations
reported here. The continuum fluxes
are in units of $10^{-15}$\,\contunits, and the line fluxes
are in units of $10^{-13}$\,\lineunits. The $\sim20$-day
time delay between continuum and emission-line variations
is apparent.}
\end{figure}

\begin{figure}
\caption{The optical continuum (upper panel) and
\Hbeta\ (lower panel) light curves from 1989 December
to 1996 November. The data are comprised of 857 continuum
measurements and 820 emission-line measurements.
The continuum fluxes
are in units of $10^{-15}$\,\contunits, and the line fluxes
are in units of $10^{-13}$\,\lineunits.}
\end{figure}

\begin{figure}
\caption{The upper panel shows 
high-state (JD 2449773, during 1995) and low-state
(JD 2448733, during 1992) Lick Observatory (set ``H'') spectra
of NGC 5548 showing extreme excursions of the
continuum brightness, and the lower panel shows the difference
spectrum obtained by subtracting the low-state spectrum from
the high-state spectrum. In the low-state spectrum, the broad
component of \Hbeta\ (just shortward of 5000\,\AA)
is very weak and the constant narrow-line component 
is thus relatively  more prominent than when the broad component
is strong. The most dramatic effect is the variability
of the ``small blue bump,'' the long-wavelength end of
which is apparent in the high-state spectrum shortward of
$\sim4200$\,\AA. }
\end{figure}

\begin{figure}
\caption{Cross-correlation functions for the continuum and
\Hbeta\ emission-line light curves of NGC 5548. The various
panels represent subsets of the data as outlined in Table 8.
The solid line shows the ICCF, and the vertical lines
represent the $1\sigma$ uncertainties associated with the
DCF values.}
\end{figure}

\begin{figure}
\caption{The upper panel shows the cross-correlation centroid
\tcent\ as a function of time, for nine years beginning with 1988
(from the Wise Observatory campaign described by Netzer et al.\ 1990,
plotted as an open circle),
followed by the eight years of International AGN Watch monitoring
(shown as filled circles).
The lower panel shows the yearly mean continuum
flux at 5100\,\AA. The error bars simply represent the rms
variations about the mean values.}
\end{figure}

\begin{figure}
\caption{The cross-correlation centroid
\tcent\ as a function the yearly mean continuum
flux at 5100\,\AA, based on eight years of International AGN Watch
data (filled circles) plus 1988 data from Wise Observatory
(open circle). The data suggest that these two quantities
may be correlated, and the slope of the best-fit line
is non-zero only at a $2.5\sigma$ confidence level.}
\end{figure}

\clearpage

%
%
\begin{deluxetable}{llc}
\tablewidth{0pt}
\tablecaption{Sources of Spectroscopic Observations}
\tablehead{
\colhead{Data} &
\colhead{ } &
\colhead{Number of} \nl
\colhead{Set} &
\colhead{Telescope and Instrument} &
\colhead{Spectra} \nl
\colhead{(1)} &
\colhead{(2)} &
\colhead{(3)} 
}
\startdata
A & 1.8-m Perkins Telescope + Ohio State CCD Spectrograph 	& 95 \nl
B & 1.0-m Wise Observatory Telescope + CCD Spectrograph   	&  4 \nl
D & 5.0-m Hale Telescope + Double Spectrograph            	&  2 \nl
F & 1.5-m Tillinghast Reflector + FAST Spectrograph       	& 89 \nl
H & 3.0-m Lick Telescope + Kast Spectrograph              	& 30 \nl
J & 2.1-m McDonald  + Cassegrain Grating  Spectrograph    	&  2 \nl
L & 6-m Special Astrophys.\ Obs.\ Telescope + CCD Spectrographs & 2 \nl
M & 3.5-m and 2.2-m Calar Alto Telescopes + CCD Spectrograph    & 2 \nl
R & 1.5-m Loiano Telescope + CCD Spectrograph                   & 2 \nl
W & 2.6-m Shajn Telescope + CCD Spectrograph              	& 76 \nl
Y & 2.2-m Beijing Astronomical Observatory + CCD Spectrograph   &  4 \nl
Z & 3.6-m ESO Telescope + CCD Spectrograph			&  1 \nl
\enddata

\end{deluxetable}

%
%
\begin{deluxetable}{lll}
\tablewidth{0pt}
\tablecaption{Absolute Calibration Check}
\tablehead{
\colhead{$F$([O\,{\sc iii}]\,$\lambda$5007)} 
	& \colhead{File} 
	& \colhead{UT} \nl
\colhead{($10^{-13}$\,ergs\,s$^{-1}$\,cm$^{-2}$)}
	& \colhead{name} 
	& \colhead{Date} \nl
\colhead{(1)} &
\colhead{(2)} &
\colhead{(3)}
}
\startdata
 5.02    \dotfill & n59326a & 1993 Dec 4  \nl
 5.09	 \dotfill & n59332a & 1993 Dec 10 \nl
 5.04	 \dotfill & n59447h & 1994 Apr 4  \nl
 4.66	 \dotfill & n59457a & 1994 Apr 15 \nl
 5.11	 \dotfill & n59464a & 1994 Apr 22 \nl
 5.22	 \dotfill & n59506h & 1994 Jun 3  \nl
 5.30	 \dotfill & n59513a & 1994 Jun 10 \nl
 5.86	 \dotfill & n59519h & 1994 Jun 13 \nl
 5.88	 \dotfill & n59548h & 1994 Jul 15 \nl
 6.37	 \dotfill & n59568h & 1994 Aug 4  \nl
 5.43	 \dotfill & n59861h & 1995 May 24 \nl
 5.49	 \dotfill & n59903a & 1995 Jul 5  \nl
 5.50	 \dotfill & n59905h & 1995 Jul 7  \nl
 5.22	 \dotfill & n59923a & 1995 Jul 25 \nl
 5.29	 \dotfill & n59958a & 1995 Aug 29  \nl
 \underline{5.75} \dotfill & \underline{n50060a} & \underline{1995 Dec 9}  \nl
 5.39 $\pm$ 0.42 \dotfill 
	& \multicolumn{2}{l}{Mean value from Years 6--8 }\nl
 5.58 $\pm$ 0.27 \dotfill 
	& \multicolumn{2}{l}{Adopted absolute flux}\nl
\enddata
\end{deluxetable}

%
%
\begin{deluxetable}{lll}
\tablewidth{0pt}
\tablecaption{Flux Scale Factors for Optical Spectra}
\tablehead{
\colhead{Data} &
\colhead{Point-Source} &
\colhead{Extended Source } \nl
\colhead{Set} &
\colhead{Scale Factor} &
\colhead{Correction $G$} \nl
\colhead{ } &
\colhead{$\varphi$} &
\colhead{($10^{-15}$ ergs s$^{-1}$\,cm$^{-2}$\,\AA$^{-1}$)} \nl
\colhead{(1)} &
\colhead{(2)} & 
\colhead{(3)} 
}
\startdata
A & $1.000$ & $0.000$ \nl
B & $0.998 \pm 0.013$ & $-0.349 \pm 0.342 $ \nl
D & $0.983 \pm 0.020$ & $-0.228 \pm 0.199 $ \nl
F(Year 6) & $0.873 \pm 0.055$ & $-1.626 \pm 0.407 $ \nl
F(Year 7) & $1.028 \pm 0.041$ & $-0.067 \pm 0.419 $ \nl
F(Year 8) & $0.965 \pm 0.035$ & $-1.344 \pm 0.373 $ \nl
H & $1.000 \pm 0.030$ & $0.000 \pm 0.267 $ \nl
J & $1.025 \pm 0.032$ & $0.046 \pm 0.471 $ \nl
L & $0.893 \pm 0.038$ & $-1.907 \pm 0.748 $ \nl
M & $0.759 \pm 0.015$ & $-1.381 \pm 0.126 $ \nl
R & $0.863 \pm 0.025$ & $-1.900 \pm 1.065 $ \nl
W & $0.940 \pm 0.040$ & $-1.149 \pm 0.388 $ \nl
Y & $1.093 \pm 0.084$ & $-0.471 \pm 0.381 $ \nl
Z & $1.001$ & $0.161 $ \nl
\enddata
\end{deluxetable}

%
%
\begin{deluxetable}{ccc}
\tablewidth{0pt}
\tablecaption{Optical Continuum and H$\beta$ Light Curves}
\tablehead{
\colhead{Julian Date} &
\colhead{$F_{\lambda}$(5100\,\AA)} &
\colhead{$F$(H$\beta$)} \nl
\colhead{$-2400000$} &
\colhead{($10^{-15}$ ergs s$^{-1}$\,cm$^{-2}$\,\AA$^{-1}$)} &
\colhead{($10^{-13}$ ergs s$^{-1}$\,cm$^{-2}$)} \nl
\colhead{(1)} &
\colhead{(2)} & 
\colhead{(3)} 
}
\startdata
 49309.1& $   8.37\pm   0.17$ & $  6.51\pm   0.13$ \nl
 49313.6& $   8.54\pm   0.28$ & $  6.94\pm   0.26$ \nl
 49317.0& $   9.32\pm   0.19$ & $  6.45\pm   0.13$ \nl
 49324.0& $   9.33\pm   0.19$ & $  6.34\pm   0.13$ \nl
 49325.0& $   9.59\pm   0.19$ & $  6.69\pm   0.13$ \nl
 49326.0& $   9.23\pm   0.19$ & $  6.73\pm   0.14$ \nl
 49327.0& $   9.14\pm   0.18$ & $  6.94\pm   0.14$ \nl
 49327.6& $   8.86\pm   0.29$ & $  6.68\pm   0.25$ \nl
 49328.0& $   9.24\pm   0.19$ & $  6.74\pm   0.14$ \nl
 49329.0& $   9.04\pm   0.18$ & $  6.86\pm   0.14$ \nl
 49330.0& $   9.27\pm   0.19$ & $  6.86\pm   0.14$ \nl
 49331.0& $   9.27\pm   0.19$ & $  6.78\pm   0.14$ \nl
 49332.0& $   9.05\pm   0.18$ & $  6.99\pm   0.14$ \nl
 49332.6& $   9.44\pm   0.31$ & $  7.02\pm   0.27$ \nl
 49333.0& $   9.30\pm   0.19$ & $  6.74\pm   0.14$ \nl
 49339.0& $   9.81\pm   0.20$ & $  6.91\pm   0.14$ \nl
 49339.7& $   9.64\pm   0.48$ & $  6.93\pm   0.35$ \nl
 49340.1& $   9.69\pm   0.19$ & $  7.15\pm   0.14$ \nl
 49342.4& $   9.78\pm   0.49$ & $  7.10\pm   0.35$ \nl
 49343.1& $   9.59\pm   0.19$ & $  7.03\pm   0.14$ \nl
 49344.1& $   9.91\pm   0.20$ & $  7.03\pm   0.14$ \nl
 49360.1& $   9.13\pm   0.18$ & $  7.12\pm   0.14$ \nl
 49364.6& $   9.02\pm   0.30$ & $  7.29\pm   0.28$ \nl
 49365.6& $   8.69\pm   0.29$ & $  7.43\pm   0.28$ \nl
 49367.0& $   8.48\pm   0.53$ & $  7.64\pm   0.43$ \nl
 49370.0& $   8.40\pm   0.52$ & $  7.33\pm   0.41$ \nl
 49371.0& $   8.74\pm   0.54$ & $  7.84\pm   0.44$ \nl
 49372.0& $   9.02\pm   0.56$ & $  7.15\pm   0.40$ \nl
 49372.1& $   9.38\pm   0.19$ & $  7.20\pm   0.14$ \nl
 49374.0& $   9.31\pm   0.58$ & $  7.26\pm   0.41$ \nl
 49375.0& $   8.98\pm   0.56$ & $  7.15\pm   0.40$ \nl
 49377.7& $   8.16\pm   0.41$ & $  7.05\pm   0.35$ \nl
 49387.0& $   7.79\pm   0.48$ & $  6.81\pm   0.38$ \nl
 49389.1& $   8.69\pm   0.17$ & $  6.83\pm   0.14$ \nl
 49390.0& $   8.00\pm   0.50$ & $  6.67\pm   0.37$ \nl
 49394.0& $   7.59\pm   0.47$ & $  6.53\pm   0.37$ \nl
 49395.0& $   8.15\pm   0.51$ & $  6.68\pm   0.37$ \nl
 49395.6& $   8.87\pm   0.29$ & $  6.53\pm   0.25$ \nl
 49397.6& $   9.11\pm   0.30$ & $  6.61\pm   0.25$ \nl
 49399.0& $   8.34\pm   0.52$ & $  6.53\pm   0.37$ \nl
 49400.0& $   7.64\pm   0.47$ & $  6.41\pm   0.36$ \nl
 49403.8& $   8.51\pm   0.17$ & $  6.49\pm   0.13$ \nl
 49409.0& $   8.51\pm   0.17$ & $  6.81\pm   0.14$ \nl
 49415.9& $   9.32\pm   0.19$ & $  6.78\pm   0.14$ \nl
 49416.9& $   9.40\pm   0.58$ & $  6.75\pm   0.38$ \nl
 49420.0& $   9.53\pm   0.59$ & $  6.58\pm   0.37$ \nl
 49420.9& $   9.45\pm   0.19$ & $  7.05\pm   0.14$ \nl
 49421.0& $   9.44\pm   0.59$ & $  6.99\pm   0.39$ \nl
 49422.0& $  10.12\pm   0.63$ & $  6.76\pm   0.38$ \nl
 49428.6& $   9.22\pm   0.30$ & $  6.88\pm   0.26$ \nl
 49429.1& $   9.19\pm   0.18$ & $  7.03\pm   0.14$ \nl
 49429.8& $   9.24\pm   0.19$ & $  7.04\pm   0.14$ \nl
 49429.9& $   8.68\pm   0.54$ & $  6.55\pm   0.37$ \nl
 49436.9& $   9.46\pm   0.19$ & $  7.17\pm   0.14$ \nl
 49443.9& $   9.90\pm   0.19$ & $  7.16\pm   0.13$ \nl
 49444.8& $   9.54\pm   0.59$ & $  7.37\pm   0.41$ \nl
 49447.0& $  10.32\pm   0.21$ & $  7.60\pm   0.15$ \nl
 49447.9& $  10.10\pm   0.63$ & $  7.35\pm   0.41$ \nl
 49448.9& $  10.56\pm   0.65$ & $  7.28\pm   0.41$ \nl
 49452.8& $  10.49\pm   0.65$ & $  7.55\pm   0.42$ \nl
 49453.3& $   9.66\pm   0.32$ & $  7.43\pm   0.28$ \nl
 49454.4& $   9.27\pm   0.31$ & $  7.57\pm   0.29$ \nl
 49457.9& $   9.41\pm   0.19$ & $  7.82\pm   0.16$ \nl
 49461.0& $   8.72\pm   0.17$ & $  7.50\pm   0.15$ \nl
 49464.8& $   8.74\pm   0.17$ & $  7.09\pm   0.14$ \nl
 49471.8& $   8.94\pm   0.18$ & $  6.67\pm   0.13$ \nl
 49475.8& $   9.54\pm   0.59$ & $  6.98\pm   0.39$ \nl
 49476.9& $  10.01\pm   0.62$ & $  7.68\pm   0.43$ \nl
 49477.8& $   9.57\pm   0.59$ & $  7.74\pm   0.43$ \nl
 49478.8& $   9.09\pm   0.18$ & $  6.26\pm   0.12$ \nl
 49481.8& $   9.64\pm   0.60$ & $  6.76\pm   0.38$ \nl
 49482.8& $   8.51\pm   0.53$ & $  6.32\pm   0.35$ \nl
 49483.7& $  10.14\pm   0.63$ & $  6.97\pm   0.39$ \nl
 49485.8& $   9.57\pm   0.19$ & $  6.88\pm   0.14$ \nl
 49486.4& $   9.25\pm   0.31$ & $  6.93\pm   0.26$ \nl
 49487.4& $   9.28\pm   0.31$ & $  6.97\pm   0.26$ \nl
 49488.8& $   8.81\pm   0.55$ & $  6.58\pm   0.37$ \nl
 49489.4& $   9.12\pm   0.30$ & $  7.13\pm   0.27$ \nl
 49489.8& $   9.22\pm   0.57$ & $  6.86\pm   0.38$ \nl
 49490.1& $   9.17\pm   0.30$ & $  7.39\pm   0.28$ \nl
 49490.4& $   9.21\pm   0.30$ & $  6.99\pm   0.27$ \nl
 49491.1& $   9.61\pm   0.32$ & $  7.17\pm   0.27$ \nl
 49491.4& $   9.33\pm   0.31$ & $  7.15\pm   0.27$ \nl
 49491.6& $   9.05\pm   0.56$ & $  7.12\pm   0.40$ \nl
 49491.8& $   9.91\pm   0.20$ & $  7.83\pm   0.16$ \nl
 49498.8& $  10.90\pm   0.22$ & $  8.09\pm   0.16$ \nl
 49504.1& $  11.16\pm   0.37$ & $  8.98\pm   0.34$ \nl
 49504.7& $  11.33\pm   0.70$ & $  8.02\pm   0.45$ \nl
 49505.6& $  11.07\pm   0.69$ & $  8.69\pm   0.49$ \nl
 49506.7& $  11.17\pm   0.21$ & $  7.99\pm   0.15$ \nl
 49506.9& $  11.47\pm   0.23$ & $  7.90\pm   0.16$ \nl
 49507.7& $  11.02\pm   0.68$ & $  8.37\pm   0.47$ \nl
 49508.7& $  12.06\pm   0.75$ & $  8.51\pm   0.48$ \nl
 49509.7& $  11.15\pm   0.69$ & $  8.30\pm   0.47$ \nl
 49510.4& $  11.10\pm   0.37$ & $  8.70\pm   0.33$ \nl
 49513.7& $  10.50\pm   0.20$ & $  8.32\pm   0.16$ \nl
 49514.7& $  10.88\pm   0.67$ & $  8.50\pm   0.48$ \nl
 49515.7& $   9.73\pm   0.60$ & $  8.29\pm   0.46$ \nl
 49518.7& $  10.03\pm   0.62$ & $  8.61\pm   0.48$ \nl
 49519.4& $   9.84\pm   0.32$ & $  8.32\pm   0.32$ \nl
 49519.5& $  10.18\pm   0.20$ & $  8.56\pm   0.17$ \nl
 49519.7& $  10.65\pm   0.66$ & $  8.57\pm   0.48$ \nl
 49520.4& $   9.54\pm   0.31$ & $  8.29\pm   0.31$ \nl
 49520.7& $  10.09\pm   0.19$ & $  7.87\pm   0.15$ \nl
 49527.7& $  11.00\pm   0.22$ & $  8.23\pm   0.17$ \nl
 49535.4& $  11.28\pm   0.37$ & $  8.44\pm   0.32$ \nl
 49536.3& $  11.69\pm   0.39$ & $  8.65\pm   0.33$ \nl
 49537.7& $  11.25\pm   0.70$ & $  8.55\pm   0.48$ \nl
 49538.7& $  11.08\pm   0.69$ & $  8.47\pm   0.47$ \nl
 49539.7& $  11.77\pm   0.73$ & $  9.02\pm   0.50$ \nl
 49545.6& $  10.97\pm   0.68$ & $  8.67\pm   0.48$ \nl
 49548.8& $  11.97\pm   0.24$ & $  8.88\pm   0.18$ \nl
 49553.4& $  13.35\pm   0.44$ & $  8.43\pm   0.32$ \nl
 49566.3& $  11.18\pm   0.37$ & $  8.99\pm   0.34$ \nl
 49568.4& $  11.64\pm   0.38$ & $  8.90\pm   0.34$ \nl
 49568.8& $  11.39\pm   0.23$ & $  9.19\pm   0.18$ \nl
 49575.5& $  11.99\pm   0.60$ & $  9.05\pm   0.45$ \nl
 49579.3& $  12.06\pm   0.40$ & $  9.11\pm   0.35$ \nl
 49597.3& $  11.72\pm   0.39$ & $  9.33\pm   0.35$ \nl
 49597.7& $  11.45\pm   0.16$ & $  9.55\pm   0.14$ \nl
 49599.3& $  11.83\pm   0.39$ & $  9.85\pm   0.37$ \nl
 49601.3& $  11.24\pm   0.37$ & $  9.82\pm   0.37$ \nl
 49603.7& $  11.18\pm   0.22$ & $  9.84\pm   0.20$ \nl
 49604.7& $  10.80\pm   0.22$ & $  9.72\pm   0.19$ \nl
 49608.7& $  11.01\pm   0.22$ & $  9.76\pm   0.19$ \nl
 49622.6& $   9.57\pm   0.19$ & $  9.45\pm   0.19$ \nl
 49626.6& $   9.06\pm   0.18$ & $  9.11\pm   0.18$ \nl
 49636.6& $   9.44\pm   0.19$ & $  8.44\pm   0.17$ \nl
 49679.0& $   9.81\pm   0.20$ & $  9.63\pm   0.19$ \nl
 49686.0& $  10.41\pm   0.21$ & $  8.82\pm   0.18$ \nl
 49713.6& $  10.82\pm   0.36$ & $  8.43\pm   0.32$ \nl
 49743.6& $  10.89\pm   0.36$ & $  8.29\pm   0.31$ \nl
 49744.5& $  10.95\pm   0.36$ & $  8.27\pm   0.31$ \nl
 49750.1& $  11.12\pm   0.22$ & $  8.24\pm   0.21$ \nl
 49751.0& $  10.77\pm   0.22$ & $  7.89\pm   0.20$ \nl
 49752.0& $  11.43\pm   0.23$ & $  8.61\pm   0.17$ \nl
 49753.6& $  12.02\pm   0.40$ & $  8.48\pm   0.32$ \nl
 49758.9& $  11.65\pm   0.23$ & $  8.48\pm   0.17$ \nl
 49765.9& $  10.34\pm   0.21$ & $  8.78\pm   0.18$ \nl
 49772.5& $  11.14\pm   0.37$ & $  8.37\pm   0.32$ \nl
 49772.9& $  11.39\pm   0.23$ & $  8.21\pm   0.16$ \nl
 49773.1& $  11.50\pm   0.23$ & $  8.40\pm   0.17$ \nl
 49775.0& $  11.68\pm   0.35$ & $  8.58\pm   0.17$ \nl
 49779.0& $  11.61\pm   0.23$ & $  8.81\pm   0.22$ \nl
 49780.0& $  11.63\pm   0.23$ & $  8.97\pm   0.22$ \nl
 49781.0& $  12.88\pm   0.26$ & $  8.94\pm   0.22$ \nl
 49782.0& $  11.92\pm   0.24$ & $  9.25\pm   0.23$ \nl
 49783.5& $  12.27\pm   0.41$ & $  9.02\pm   0.34$ \nl
 49784.0& $  12.47\pm   0.25$ & $  9.08\pm   0.23$ \nl
 49784.5& $  12.16\pm   0.24$ & $  8.99\pm   0.18$ \nl
 49784.9& $  12.00\pm   0.24$ & $  8.89\pm   0.18$ \nl
 49786.6& $  12.40\pm   0.25$ & $  9.09\pm   0.18$ \nl
 49787.6& $  12.17\pm   0.24$ & $  8.97\pm   0.18$ \nl
 49788.5& $  12.39\pm   0.25$ & $  8.90\pm   0.18$ \nl
 49794.0& $  12.32\pm   0.25$ & $  8.63\pm   0.17$ \nl
 49802.0& $  13.31\pm   0.27$ & $  9.14\pm   0.18$ \nl
 49808.9& $  13.50\pm   0.27$ & $  9.46\pm   0.19$ \nl
 49810.8& $  14.25\pm   0.28$ & $  9.82\pm   0.25$ \nl
 49810.9& $  14.00\pm   0.42$ & $  9.82\pm   0.20$ \nl
 49811.8& $  14.17\pm   0.28$ & $ 10.10\pm   0.25$ \nl
 49813.0& $  13.88\pm   0.28$ & $  9.72\pm   0.24$ \nl
 49813.6& $  14.14\pm   0.47$ & $  9.74\pm   0.37$ \nl
 49818.8& $  14.56\pm   0.29$ & $  9.85\pm   0.20$ \nl
 49830.0& $  14.13\pm   0.28$ & $ 10.43\pm   0.21$ \nl
 49833.8& $  13.96\pm   0.28$ & $ 10.41\pm   0.21$ \nl
 49839.8& $  13.77\pm   0.28$ & $ 10.40\pm   0.21$ \nl
 49845.4& $  13.42\pm   0.44$ & $ 10.92\pm   0.41$ \nl
 49846.8& $  13.11\pm   0.26$ & $ 10.07\pm   0.20$ \nl
 49853.8& $  12.47\pm   0.25$ & $ 10.87\pm   0.22$ \nl
 49860.8& $  11.88\pm   0.24$ & $ 10.64\pm   0.21$ \nl
 49861.9& $  11.96\pm   0.24$ & $ 10.16\pm   0.20$ \nl
 49862.3& $  12.17\pm   0.40$ & $ 10.29\pm   0.39$ \nl
 49863.3& $  11.97\pm   0.40$ & $ 10.39\pm   0.40$ \nl
 49868.8& $  12.19\pm   0.24$ & $ 10.09\pm   0.20$ \nl
 49870.4& $  12.11\pm   0.40$ & $  9.94\pm   0.38$ \nl
 49871.4& $  12.34\pm   0.41$ & $ 10.03\pm   0.38$ \nl
 49871.7& $  12.28\pm   0.25$ & $ 10.03\pm   0.25$ \nl
 49874.7& $  12.03\pm   0.24$ & $  9.76\pm   0.19$ \nl
 49875.4& $  12.06\pm   0.40$ & $  9.54\pm   0.36$ \nl
 49878.4& $  12.82\pm   0.42$ & $  9.77\pm   0.37$ \nl
 49881.7& $  11.31\pm   0.23$ & $  9.73\pm   0.19$ \nl
 49889.4& $  12.15\pm   0.40$ & $ 10.27\pm   0.39$ \nl
 49889.7& $  11.56\pm   0.23$ & $  9.26\pm   0.19$ \nl
 49891.4& $  11.71\pm   0.39$ & $  9.51\pm   0.36$ \nl
 49895.7& $  11.62\pm   0.23$ & $  9.04\pm   0.18$ \nl
 49903.3& $  11.64\pm   0.38$ & $  9.52\pm   0.36$ \nl
 49903.7& $  11.48\pm   0.23$ & $  8.98\pm   0.18$ \nl
 49905.8& $  12.08\pm   0.24$ & $  9.05\pm   0.18$ \nl
 49910.7& $  11.86\pm   0.24$ & $  8.64\pm   0.17$ \nl
 49917.7& $  11.69\pm   0.23$ & $  8.62\pm   0.17$ \nl
 49923.7& $  12.21\pm   0.24$ & $  8.83\pm   0.18$ \nl
 49923.8& $  11.95\pm   0.24$ & $  9.33\pm   0.19$ \nl
 49930.8& $  12.27\pm   0.25$ & $  9.28\pm   0.19$ \nl
 49935.3& $  11.80\pm   0.39$ & $  9.38\pm   0.36$ \nl
 49937.7& $  11.67\pm   0.23$ & $  9.58\pm   0.19$ \nl
 49953.3& $  11.07\pm   0.37$ & $  8.85\pm   0.34$ \nl
 49953.7& $  11.31\pm   0.23$ & $  9.10\pm   0.18$ \nl
 49954.7& $  11.01\pm   0.22$ & $  9.04\pm   0.18$ \nl
 49958.7& $  10.91\pm   0.22$ & $  9.18\pm   0.18$ \nl
 49966.6& $  10.73\pm   0.22$ & $  8.63\pm   0.17$ \nl
 49972.6& $  10.85\pm   0.22$ & $  8.46\pm   0.17$ \nl
 49980.3& $  12.59\pm   0.41$ & $  8.31\pm   0.32$ \nl
 49980.6& $  12.12\pm   0.24$ & $  8.69\pm   0.17$ \nl
 49985.6& $  12.51\pm   0.25$ & $  8.90\pm   0.18$ \nl
 49986.6& $  12.81\pm   0.26$ & $  8.74\pm   0.17$ \nl
 50008.2& $  12.83\pm   0.42$ & $  9.54\pm   0.36$ \nl
 50044.0& $  11.60\pm   0.23$ & $  9.66\pm   0.19$ \nl
 50048.6& $  11.22\pm   0.37$ & $  9.17\pm   0.35$ \nl
 50053.0& $  10.40\pm   0.21$ & $  9.11\pm   0.18$ \nl
 50061.0& $   8.64\pm   0.17$ & $  9.27\pm   0.19$ \nl
 50064.6& $   8.37\pm   0.28$ & $  8.67\pm   0.33$ \nl
 50074.0& $   8.05\pm   0.31$ & $  7.56\pm   0.21$ \nl
 50075.0& $   8.18\pm   0.31$ & $  7.35\pm   0.21$ \nl
 50078.0& $   7.99\pm   0.30$ & $  7.18\pm   0.20$ \nl
 50080.0& $   7.52\pm   0.29$ & $  7.14\pm   0.20$ \nl
 50081.0& $   8.25\pm   0.31$ & $  6.80\pm   0.19$ \nl
 50087.0& $   8.21\pm   0.16$ & $  6.74\pm   0.14$ \nl
 50094.0& $   8.40\pm   0.17$ & $  6.40\pm   0.13$ \nl
 50095.0& $   7.88\pm   0.30$ & $  6.40\pm   0.18$ \nl
 50096.0& $   8.10\pm   0.31$ & $  6.08\pm   0.17$ \nl
 50097.0& $   8.18\pm   0.31$ & $  6.63\pm   0.19$ \nl
 50098.1& $   7.76\pm   0.29$ & $  6.82\pm   0.19$ \nl
 50101.9& $   8.60\pm   0.17$ & $  6.35\pm   0.13$ \nl
 50102.0& $   8.90\pm   0.34$ & $  6.78\pm   0.19$ \nl
 50104.0& $   9.39\pm   0.36$ & $  6.72\pm   0.19$ \nl
 50105.0& $   9.73\pm   0.37$ & $  6.30\pm   0.18$ \nl
 50107.0& $   9.41\pm   0.36$ & $  6.75\pm   0.19$ \nl
 50108.0& $   9.50\pm   0.36$ & $  6.75\pm   0.19$ \nl
 50109.0& $   9.36\pm   0.36$ & $  6.64\pm   0.19$ \nl
 50110.0& $   9.34\pm   0.35$ & $  6.59\pm   0.19$ \nl
 50111.0& $   9.82\pm   0.37$ & $  6.81\pm   0.19$ \nl
 50112.0& $   9.34\pm   0.35$ & $  6.94\pm   0.19$ \nl
 50119.0& $  10.06\pm   0.20$ & $  7.49\pm   0.15$ \nl
 50122.9& $  10.04\pm   0.20$ & $  7.67\pm   0.15$ \nl
 50124.0& $  10.13\pm   0.38$ & $  7.36\pm   0.21$ \nl
 50124.9& $  10.17\pm   0.39$ & $  7.59\pm   0.21$ \nl
 50126.0& $  10.27\pm   0.39$ & $  7.86\pm   0.22$ \nl
 50127.6& $   9.80\pm   0.22$ & $  8.07\pm   0.16$ \nl
 50128.0& $  10.02\pm   0.20$ & $  7.94\pm   0.16$ \nl
 50128.4& $   9.81\pm   0.22$ & $  7.82\pm   0.16$ \nl
 50129.0& $   9.56\pm   0.36$ & $  7.82\pm   0.22$ \nl
 50130.0& $   9.44\pm   0.36$ & $  8.29\pm   0.23$ \nl
 50131.0& $   9.78\pm   0.37$ & $  8.20\pm   0.23$ \nl
 50132.0& $   9.68\pm   0.37$ & $  8.33\pm   0.23$ \nl
 50133.0& $  10.47\pm   0.40$ & $  8.38\pm   0.24$ \nl
 50134.0& $   9.02\pm   0.34$ & $  7.77\pm   0.22$ \nl
 50134.9& $   8.82\pm   0.34$ & $  7.96\pm   0.22$ \nl
 50135.5& $   9.10\pm   0.30$ & $  7.96\pm   0.30$ \nl
 50136.9& $   8.86\pm   0.18$ & $  8.12\pm   0.16$ \nl
 50137.0& $   9.14\pm   0.35$ & $  8.51\pm   0.24$ \nl
 50142.9& $   8.43\pm   0.17$ & $  8.05\pm   0.16$ \nl
 50149.9& $   8.45\pm   0.17$ & $  7.22\pm   0.14$ \nl
 50153.6& $  10.01\pm   0.50$ & $  7.20\pm   0.36$ \nl
 50155.5& $   9.68\pm   0.32$ & $  6.86\pm   0.26$ \nl
 50156.5& $   9.96\pm   0.33$ & $  7.20\pm   0.27$ \nl
 50161.5& $  10.27\pm   0.34$ & $  7.23\pm   0.28$ \nl
 50164.8& $  10.18\pm   0.20$ & $  7.46\pm   0.15$ \nl
 50168.9& $   9.88\pm   0.20$ & $  7.75\pm   0.16$ \nl
 50169.0& $  10.54\pm   0.40$ & $  7.59\pm   0.21$ \nl
 50170.0& $  10.39\pm   0.40$ & $  7.61\pm   0.21$ \nl
 50175.9& $  10.86\pm   0.22$ & $  8.12\pm   0.16$ \nl
 50185.9& $  12.54\pm   0.25$ & $  8.02\pm   0.16$ \nl
 50188.4& $  12.30\pm   0.41$ & $  8.25\pm   0.31$ \nl
 50191.8& $  12.07\pm   0.24$ & $  8.55\pm   0.17$ \nl
 50198.8& $  11.22\pm   0.22$ & $  9.03\pm   0.18$ \nl
 50200.4& $  10.80\pm   0.36$ & $  9.16\pm   0.35$ \nl
 50201.3& $  10.49\pm   0.35$ & $  9.06\pm   0.34$ \nl
 50206.9& $  10.90\pm   0.22$ & $  9.04\pm   0.18$ \nl
 50212.5& $  10.66\pm   0.35$ & $  9.15\pm   0.35$ \nl
 50212.9& $  10.31\pm   0.21$ & $  8.71\pm   0.17$ \nl
 50213.4& $  10.37\pm   0.34$ & $  9.20\pm   0.35$ \nl
 50215.4& $  10.23\pm   0.34$ & $  9.14\pm   0.35$ \nl
 50220.8& $   9.21\pm   0.18$ & $  8.73\pm   0.17$ \nl
 50224.4& $   9.48\pm   0.31$ & $  8.90\pm   0.34$ \nl
 50226.4& $  10.01\pm   0.33$ & $  8.78\pm   0.33$ \nl
 50227.8& $   9.17\pm   0.18$ & $  8.39\pm   0.17$ \nl
 50233.8& $   9.40\pm   0.19$ & $  7.94\pm   0.16$ \nl
 50241.8& $   9.48\pm   0.19$ & $  7.50\pm   0.15$ \nl
 50245.4& $   9.92\pm   0.33$ & $  7.67\pm   0.29$ \nl
 50247.4& $  10.18\pm   0.34$ & $  7.94\pm   0.30$ \nl
 50248.8& $  10.34\pm   0.21$ & $  7.36\pm   0.15$ \nl
 50253.8& $  10.89\pm   0.22$ & $  7.43\pm   0.15$ \nl
 50256.5& $  11.29\pm   0.37$ & $  7.86\pm   0.30$ \nl
 50256.9& $  11.46\pm   0.23$ & $  7.63\pm   0.15$ \nl
 50258.4& $  12.04\pm   0.40$ & $  7.80\pm   0.30$ \nl
 50262.8& $  12.04\pm   0.24$ & $  8.00\pm   0.16$ \nl
 50273.8& $  12.24\pm   0.25$ & $  9.01\pm   0.18$ \nl
 50276.4& $  13.82\pm   0.46$ & $  8.58\pm   0.33$ \nl
 50277.4& $  12.97\pm   0.43$ & $  8.84\pm   0.34$ \nl
 50280.7& $  13.17\pm   0.26$ & $  7.82\pm   0.16$ \nl
 50284.4& $  13.43\pm   0.44$ & $  8.93\pm   0.34$ \nl
 50285.4& $  13.52\pm   0.45$ & $  9.27\pm   0.35$ \nl
 50286.4& $  13.35\pm   0.44$ & $  9.10\pm   0.35$ \nl
 50305.7& $  12.11\pm   0.24$ & $ 10.03\pm   0.20$ \nl
 50315.3& $  12.51\pm   0.41$ & $  9.37\pm   0.36$ \nl
 50336.6& $  13.36\pm   0.27$ & $  9.00\pm   0.18$ \nl
 50339.2& $  13.75\pm   0.45$ & $  9.46\pm   0.36$ \nl
 50344.2& $  14.19\pm   0.47$ & $  9.59\pm   0.36$ \nl
 50350.6& $  14.40\pm   0.29$ & $  8.37\pm   0.17$ \nl
 50358.6& $  13.93\pm   0.28$ & $  9.44\pm   0.19$ \nl
 50361.6& $  13.92\pm   0.28$ & $  9.46\pm   0.19$ \nl
 50362.2& $  14.79\pm   0.49$ & $  9.35\pm   0.35$ \nl
 50372.2& $  14.48\pm   0.48$ & $  9.39\pm   0.36$ \nl
 50373.2& $  14.23\pm   0.47$ & $  9.23\pm   0.35$ \nl
\enddata
\end{deluxetable}

%
%
\begin{deluxetable}{lccc}
\tablewidth{0pt}
\tablecaption{$V$-Band CCD Photometry}
\tablehead{
\colhead{Julian Date} &
\colhead{$V$} &
\colhead{ } &
\colhead{$F_{\lambda}$(5100\,\AA)} \nl
\colhead{(-2400000)} &
\colhead{(mag)} &
\colhead{Source} &
\colhead{($10^{-15}$\,ergs s$^{-1}$\,cm$^{-2}$\,\AA$^{-1}$)} \nl
\colhead{(1)} &
\colhead{(2)} & 
\colhead{(3)} &
\colhead{(4)} 
}
\startdata
 49458.2 &$ 13.958 \pm 0.050$ & A & $9.90 \pm 0.46$ \nl
 49459.2 &$ 13.955 \pm 0.035$ & A & $9.90 \pm 0.36$ \nl
 49460.2 &$ 14.082 \pm 0.050$ & A & $7.92 \pm 0.37$ \nl
 49460.9 &$ 13.807 \pm 0.170$ & B & $8.78 \pm 1.37$ \nl
 49461.2 &$ 14.058 \pm 0.035$ & A & $8.36 \pm 0.31$ \nl
 49462.2 &$ 14.079 \pm 0.029$ & A & $7.93 \pm 0.22$ \nl
 49463.9 &$ 13.779 \pm 0.073$ & B & $9.23 \pm 0.59$ \nl
 49703.1 &$ 13.695 \pm 0.042$ & C & $10.51 \pm 0.39$ \nl
 49751.0 &$ 13.661 \pm 0.068$ & C & $11.24 \pm 0.72$ \nl
 49803.9 &$ 13.548 \pm 0.064$ & C & $13.21 \pm 0.73$ \nl
 49851.8 &$ 13.513 \pm 0.024$ & C & $14.00 \pm 0.26$ \nl
 49877.8 &$ 13.584 \pm 0.046$ & C & $12.65 \pm 0.58$ \nl
 49911.8 &$ 13.682 \pm 0.030$ & C & $10.86 \pm 0.30$ \nl
 49952.6 &$ 13.694 \pm 0.031$ & C & $10.51 \pm 0.29$ \nl
\hline
\multicolumn{4}{c}{Codes for Data Origin }\nl
\multicolumn{1}{r}{A} &
\multicolumn{3}{l}{Yunnan Astronomical Observatory, Kunming, China}\nl
\multicolumn{1}{r}{B} & 
\multicolumn{3}{l}{1.0-m Lick Telescope} \nl
\multicolumn{1}{r}{C} & 
\multicolumn{3}{l}{0.61-m UCLA Telescope}\nl
\enddata
\end{deluxetable}

%
%
\begin{deluxetable}{lcccccc}
\tablewidth{0pt}
\tablecaption{Sampling Statistics for Optical Continuum}
\tablehead{
\colhead{ } &
\colhead{Number of} &
\multicolumn{2}{c}{Sampling Interval (days)} &
\colhead{Mean} \nl
\colhead{Subset} &
\colhead{Epochs} &
\colhead{Average} &
\colhead{Median} &
\colhead{Flux\tablenotemark{a}} &
\colhead{$F_{\rm var}$} &
\colhead{$R_{\rm max}$} \nl
\colhead{(1)} &
\colhead{(2)} & 
\colhead{(3)} & 
\colhead{(4)} & 
\colhead{(5)} & 
\colhead{(6)} & 
\colhead{(7)} 
}
\startdata
All data      & 857 & 3.3 & 1.0 & $9.35\pm1.86$ & 0.195&$ 2.98\pm0.16$ \nl
Year 1 (1989) & 125 & 2.4 & 1.0 & $9.92\pm1.26$ & 0.117&$ 2.16\pm0.16$ \nl
Year 2 (1990) & 94  & 3.4 & 2.0 & $7.25\pm1.00$ & 0.129&$ 1.82\pm0.09$ \nl
Year 3 (1991) & 65  & 4.8 & 3.0 & $9.40\pm0.93$ & 0.090&$ 1.51\pm0.09$ \nl
Year 4 (1992) & 83  & 3.4 & 2.0 & $6.72\pm1.17$ & 0.168&$ 2.04\pm0.10$ \nl
Year 5 (1993) & 174 & 1.3 & 0.7 & $9.04\pm0.90$ & 0.092&$ 1.65\pm0.08$ \nl
Year 6 (1994) & 135 & 2.4 & 1.0 & $9.76\pm1.10$ & 0.104&$ 1.76\pm0.12$ \nl
Year 7 (1995) & 83  & 4.0 & 1.9 & $    12.09\pm1.00$& 0.079&$ 1.48\pm0.04$ \nl
Year 8 (1996) & 98  & 3.4 & 2.0 & $    10.47\pm1.82$& 0.171&$ 1.97\pm0.10$ \nl
\enddata
\tablenotetext{a}{Units of $10^{-15}$ ergs s$^{-1}$\,cm$^{-2}$\,\AA$^{-1}$.} 
\end{deluxetable}

%
%
\begin{deluxetable}{lcccccc}
\tablewidth{0pt}
\tablecaption{Sampling Statistics for H$\beta$ Emission Line}
\tablehead{
\colhead{ } &
\colhead{Number of} &
\multicolumn{2}{c}{Sampling Interval (days)} &
\colhead{Mean} \nl
\colhead{Subset} &
\colhead{Epochs} &
\colhead{Average} &
\colhead{Median} &
\colhead{Flux\tablenotemark{a}} &
\colhead{$F_{\rm var}$} &
\colhead{$R_{\rm max}$} \nl
\colhead{(1)} &
\colhead{(2)} & 
\colhead{(3)} & 
\colhead{(4)} & 
\colhead{(5)} & 
\colhead{(6)} & 
\colhead{(7)} 
}
\startdata
All data      & 820 & 3.5 & 1.4 & $ 7.56\pm1.53$ & 0.199&$ 4.17\pm0.30$ \nl
Year 1 (1989) & 132 & 2.3 & 1.0 & $ 8.62\pm0.85$ & 0.091&$ 1.57\pm0.12$ \nl
Year 2 (1990) & 94  & 3.4 & 2.0 & $ 5.98\pm1.17$ & 0.191&$ 2.30\pm0.12$ \nl
Year 3 (1991) & 65  & 4.8 & 3.0 & $ 7.46\pm0.81$ & 0.093&$ 1.58\pm0.14$ \nl
Year 4 (1992) & 83  & 3.4 & 2.0 & $ 4.96\pm1.44$ & 0.284&$ 3.03\pm0.30$ \nl
Year 5 (1993) & 142 & 2.1 & 1.0 & $ 7.93\pm0.53$ & 0.057&$ 1.40\pm0.06$ \nl
Year 6 (1994) & 128 & 2.6 & 1.0 & $ 7.58\pm0.94$ & 0.117&$ 1.57\pm0.07$ \nl
Year 7 (1995) & 78  & 4.2 & 2.1 & $ 9.27\pm0.70$ & 0.071&$ 1.38\pm0.06$ \nl
Year 8 (1996) & 98  & 3.4 & 2.0 & $ 8.03\pm0.96$ & 0.116&$ 1.65\pm0.06$ \nl
\enddata
\tablenotetext{a}{Units of $10^{-13}$ ergs s$^{-1}$\,cm$^{-2}$.} 
\end{deluxetable}

%
%
\begin{deluxetable}{lcccc}
\tablewidth{0pt}
\tablecaption{Cross-Correlation Results}
\tablehead{
\colhead{ } &
\colhead{$\tau_{\rm cent}$} &
\colhead{$\tau_{\rm peak}$} &
\colhead{ } &
\colhead{FWHM} \nl
\colhead{Subset} &
\colhead{(days)} &
\colhead{(days)} &
\colhead{$r_{\rm max}$} &
\colhead{(days)} \nl
\colhead{(1)} &
\colhead{(2)} & 
\colhead{(3)} & 
\colhead{(4)} & 
\colhead{(5)}  
}
\startdata
All data     & $21.57^{+2.41}_{-0.68}$& $16.5^{+2.1}_{-1.1}$& 0.896 & 199\nl
Year 1 (1989)& $19.73^{+2.03}_{-1.40}$& $21.5^{+3.0}_{-4.1}$& 0.877 & 39 \nl
Year 2 (1990)& $19.34^{+1.86}_{-2.96}$& $18.5^{+2.0}_{-0.7}$& 0.909 & 49 \nl
Year 3 (1991)& $16.35^{+3.75}_{-3.28}$& $17.5^{+2.7}_{-5.7}$& 0.740 & 33 \nl
Year 4 (1992)& $11.37^{+2.30}_{-2.30}$& $13.7^{+0.6}_{-4.7}$& 0.918 & 63 \nl
Year 5 (1993)& $13.61^{+1.41}_{-1.91}$& $13.2^{+1.9}_{-3.3}$& 0.730 & 29 \nl
Year 6 (1994)& $15.49^{+2.26}_{-6.09}$
                            & $\phantom{1}8.7^{+8.5}_{-2.4}$& 0.832 & 129\nl
Year 7 (1995)& $21.43^{+2.31}_{-3.00}$& $22.9^{+5.1}_{-3.1}$& 0.880 & 60 \nl
Year 8 (1996)& $16.85^{+1.45}_{-1.37}$& $16.3^{+1.1}_{-1.5}$& 0.921 & 51 \nl
\enddata
\end{deluxetable}

\clearpage

\begin{references}
\reference{}Alloin, D., Clavel, J., Peterson, B.M., Reichert, G.A., \& 
Stirpe, G.M. 1994, in Frontiers of Space and Ground-Based Astronomy, ed.\ 
W. Wamsteker, M.S. Longair, \& Y. Kondo  (Dordrecht: Kluwer), p.\ 423
\reference{}Baldwin, J., Ferland, G., Korista, K., \& Verner, D. 1995,
ApJ, 455, L119
\reference{}Blandford, R.D., \& McKee, C.F. 1982, ApJ, 255, 419
\reference{}Clavel, J., et al. 1991, ApJ, 366, 64
\reference{}Edelson, R.A., \& Krolik, J.H. 1988, ApJ, 333, 646
\reference{}Fabricant, D., Cheimets, P., Caldwell, N., \&
Geary, J. 1998, PASP, 110, 79
\reference{}Gaskell, C.M., \& Peterson, B.M. 1987, ApJS, 65, 1
\reference{}Gaskell, C.M., \& Sparke, L.S. 1986, ApJ, 305, 175
\reference{}Koratkar, A.P., \& Gaskell, C.M. 1991, ApJS, 75, 719
\reference{}Korista, K.T., et al. 1995, ApJS, 97, 285
\reference{}Kraemer, S.B., Crenshaw, D.M., Filippenko, A.V., \&
Peterson, B.M. 1998, ApJ, 498, 719
\reference{}Maoz, D., et al. 1993, ApJ, 404, 576 
\reference{}Marshall, H.L., et al. 1997,  ApJ, 479, 222
\reference{}Netzer, H., \& Maoz, D. 1990, ApJ, 365, L5
\reference{}Netzer, H., Maoz, D., Laor, A., Mendelson, H., Brosch, N.,
Leibowitz, E., Almoznino, E., Beck, S., \& Mazeh, T. 1990, ApJ, 353, 108
\reference{}Netzer, H., \& Peterson, B.M. 1997, in Astronomical Time Series,
ed.\ D.\ Maoz, A.\ Sternberg, \& E.M.\ Leibowitz (Dordrecht: Kluwer), p.\ 85
\reference{}Penston, M.V. 1991, in Variability of Galactic Nuclei,
ed.\ H.R.\ Miller \& P.J.\ Wiita (Cambridge: Cambridge Univ.\ Press),
p.\ 343
\reference{}Peterson, B.M. 1993, PASP, 105, 247
\reference{}Peterson, B.M., et al. 1991, ApJ, 368, 119
\reference{}Peterson, B.M., et al. 1992, ApJ, 392, 470
\reference{}Peterson, B.M., et al. 1994, ApJ,  425, 622
\reference{}Peterson, B.M., Pogge, R.W., Wanders, I., Smith, S.M.,
\& Romanishin, W. 1995, PASP, 107, 579
\reference{}Peterson, B.M., Wanders, I., Bertram, R.,
Hunley, J.F., Pogge, R.W., \& Wagner, R.M. 1998a, ApJ, 501, in press
\reference{}Peterson, B.M., Wanders, I., Horne, K., Collier, S.,
Alexander, T., Kaspi, S., \& Maoz, D. 1998b, PASP, 110, 660
\reference{}Rodr\'{\i}guez, P.M., et al. 1997, ApJ, 110, 9
\reference{}Romanishin, W., et al.\ 1995, ApJ, 455, 516
\reference{}Sergeev, S.G., Pronik, V.I., Malkov, Yu.F.,
\& Chuvaev, K.K. 1997, A\&A, 320, 405
\reference{}Ulrich, M.-H., Boksenberg, A., Bromage, G.E., 
Clavel, J., Elvius, A., Penston, M.V., Perola, G.C., \&
Snijders, M.A.J. 1991, ApJ, 382, 483
\reference{}van Groningen, E., \& Wanders, I. 1992, PASP, 104, 700
\reference{}Wanders, I. 1995, A\&A, 296, 332
\reference{}Wanders, I., \& Peterson, B.M. 1996, ApJ, 466, 174
\reference{}Wanders, I., et al., 1997, ApJS, 113, 69
\reference{}White, R.J., \& Peterson, B.M. 1994, PASP, 106, 879
\end{references}
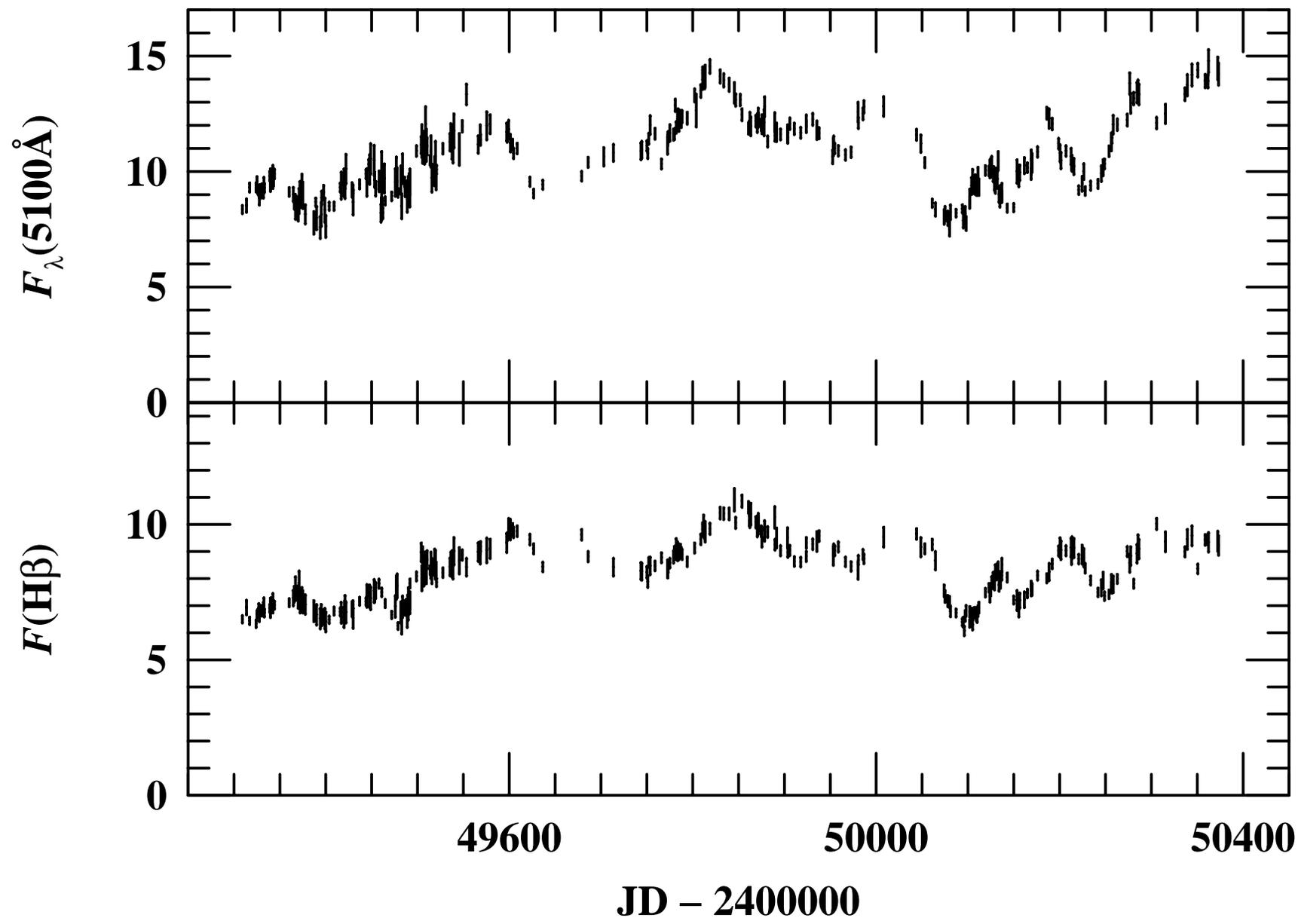
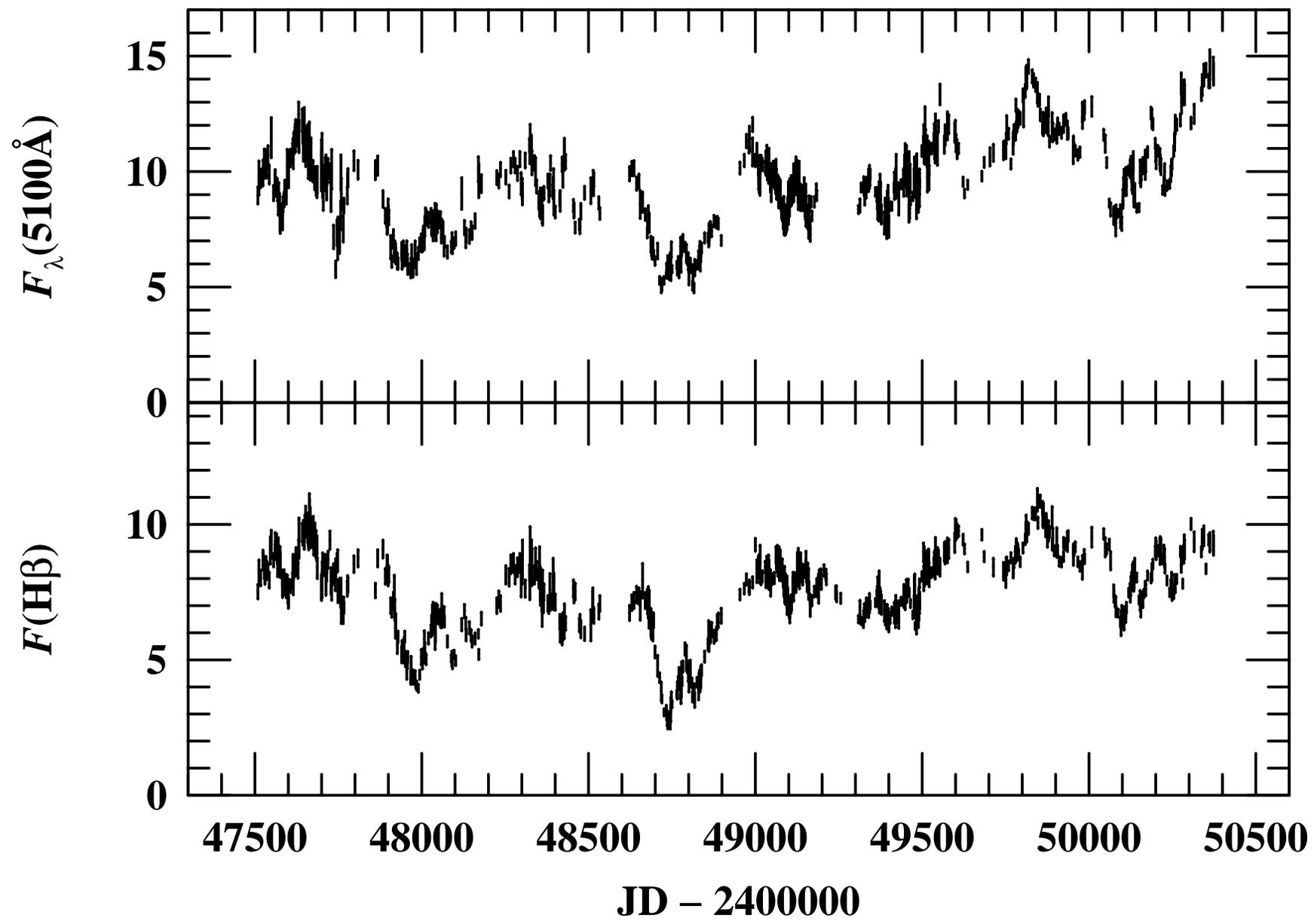
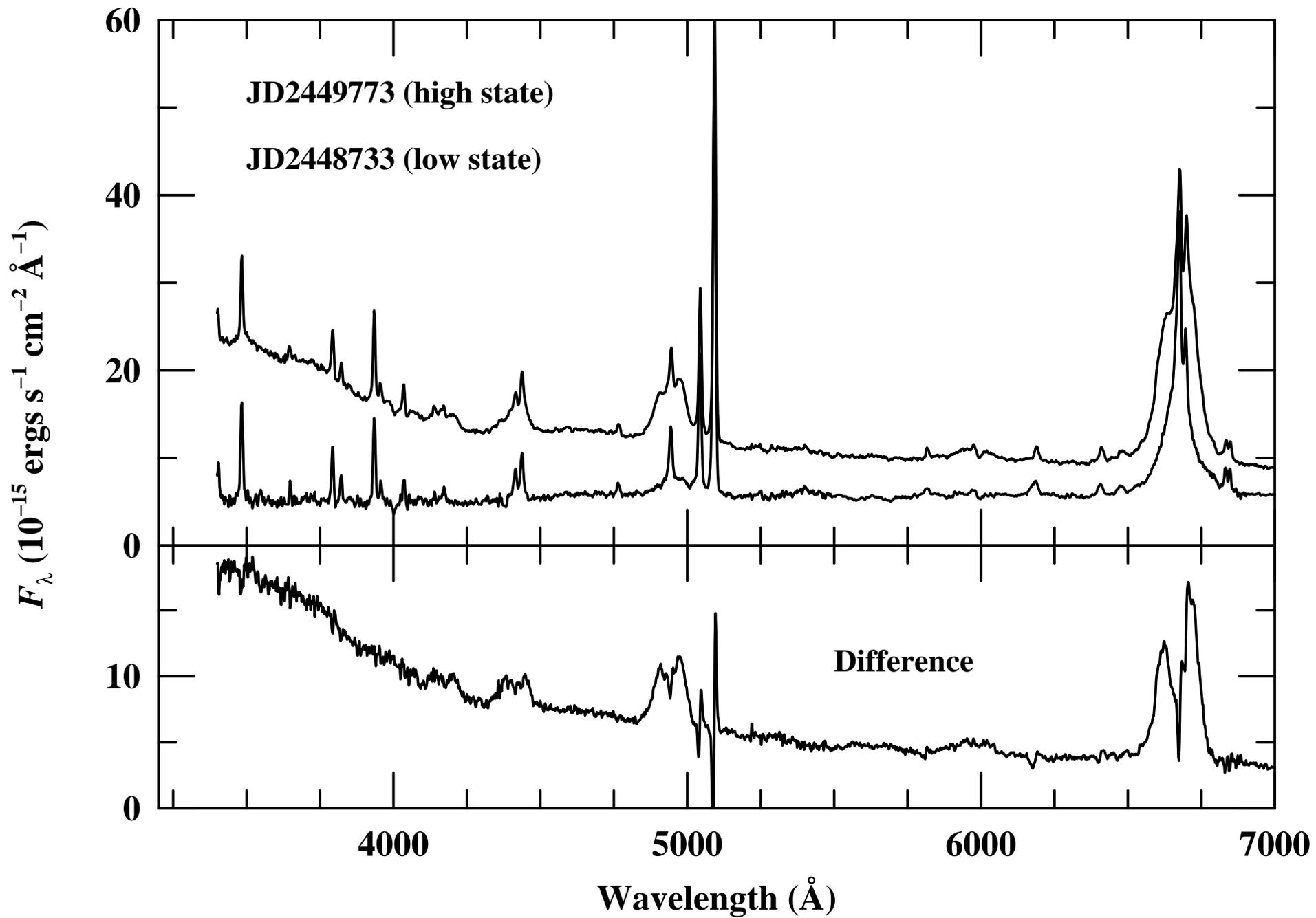
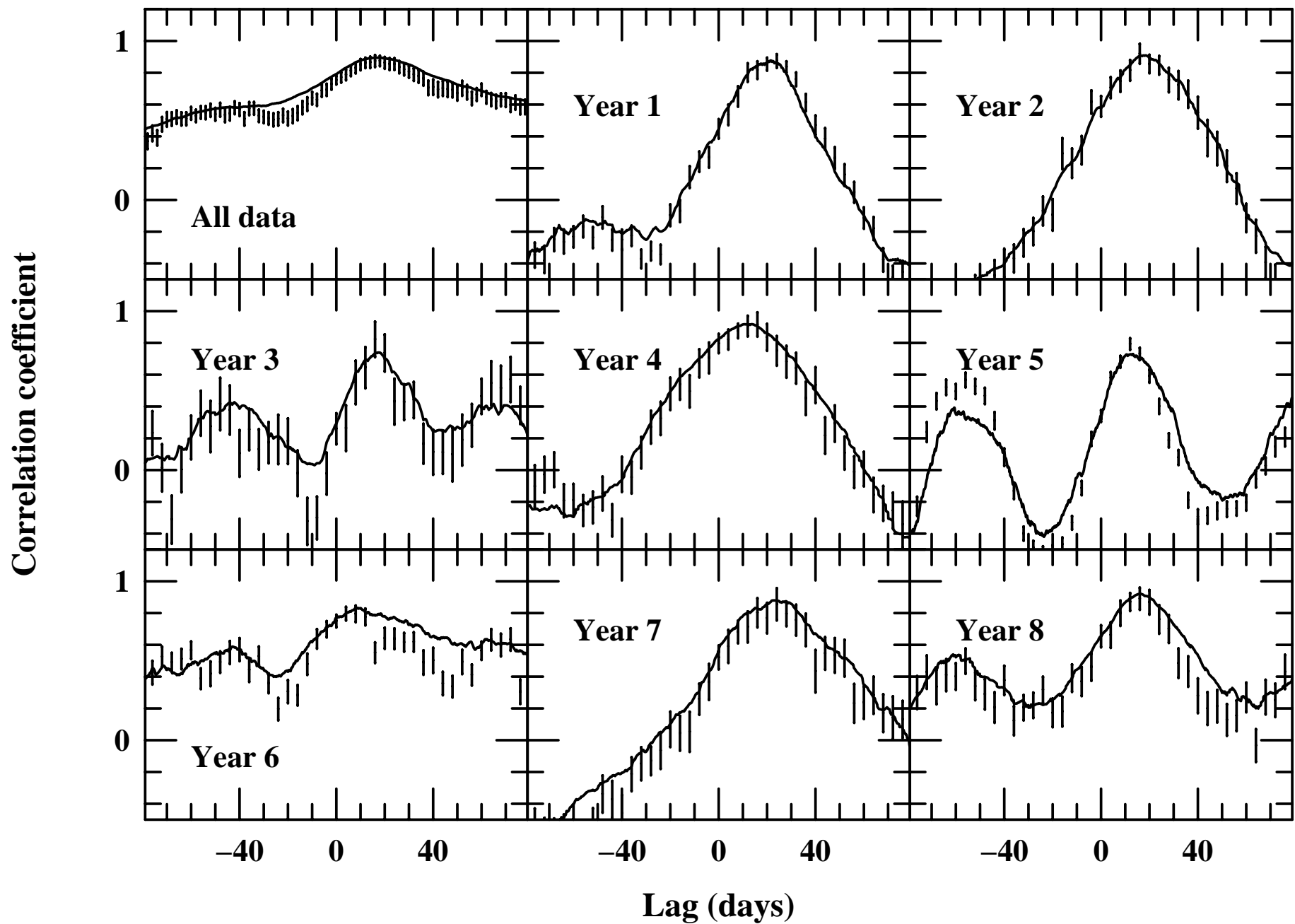
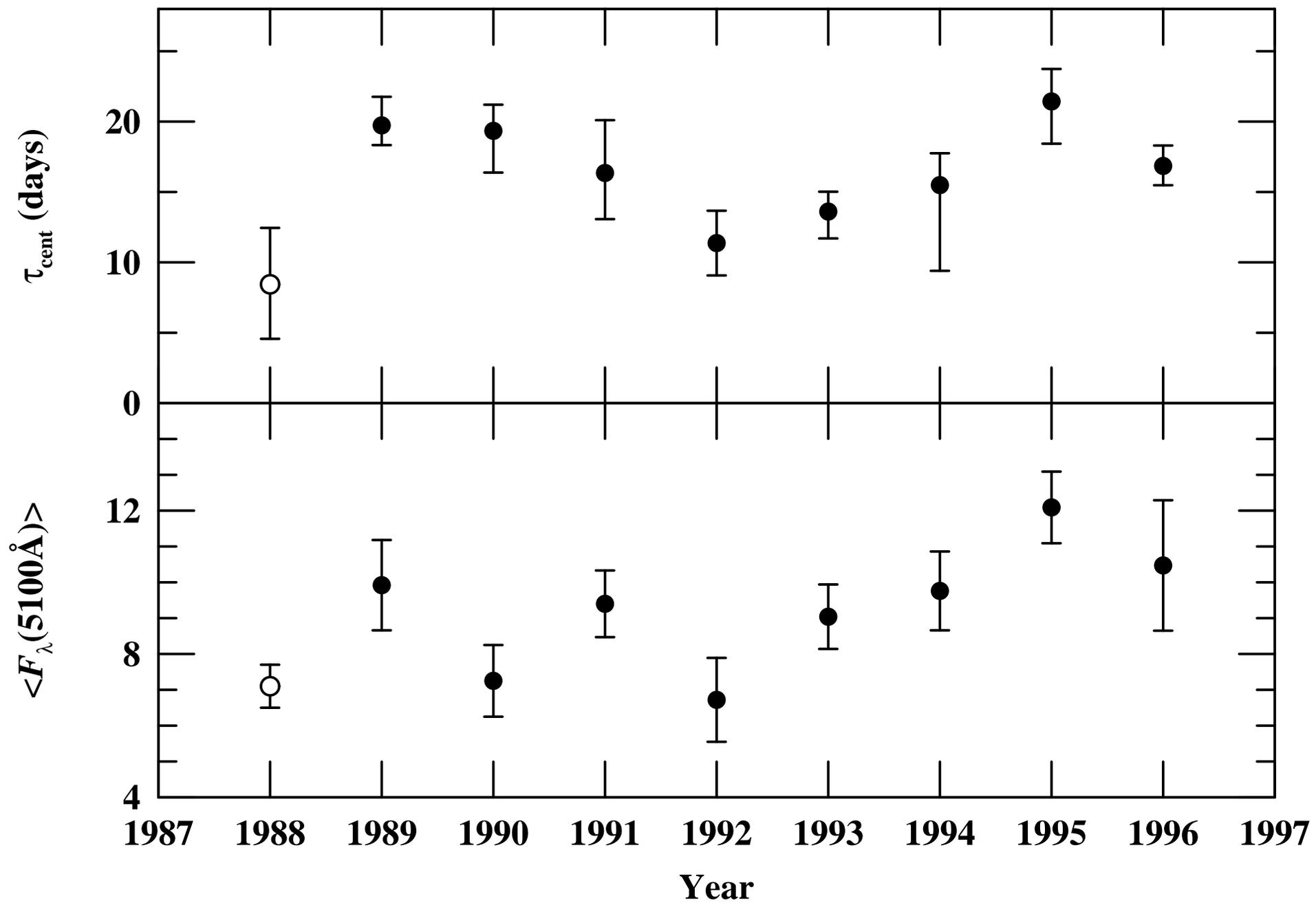
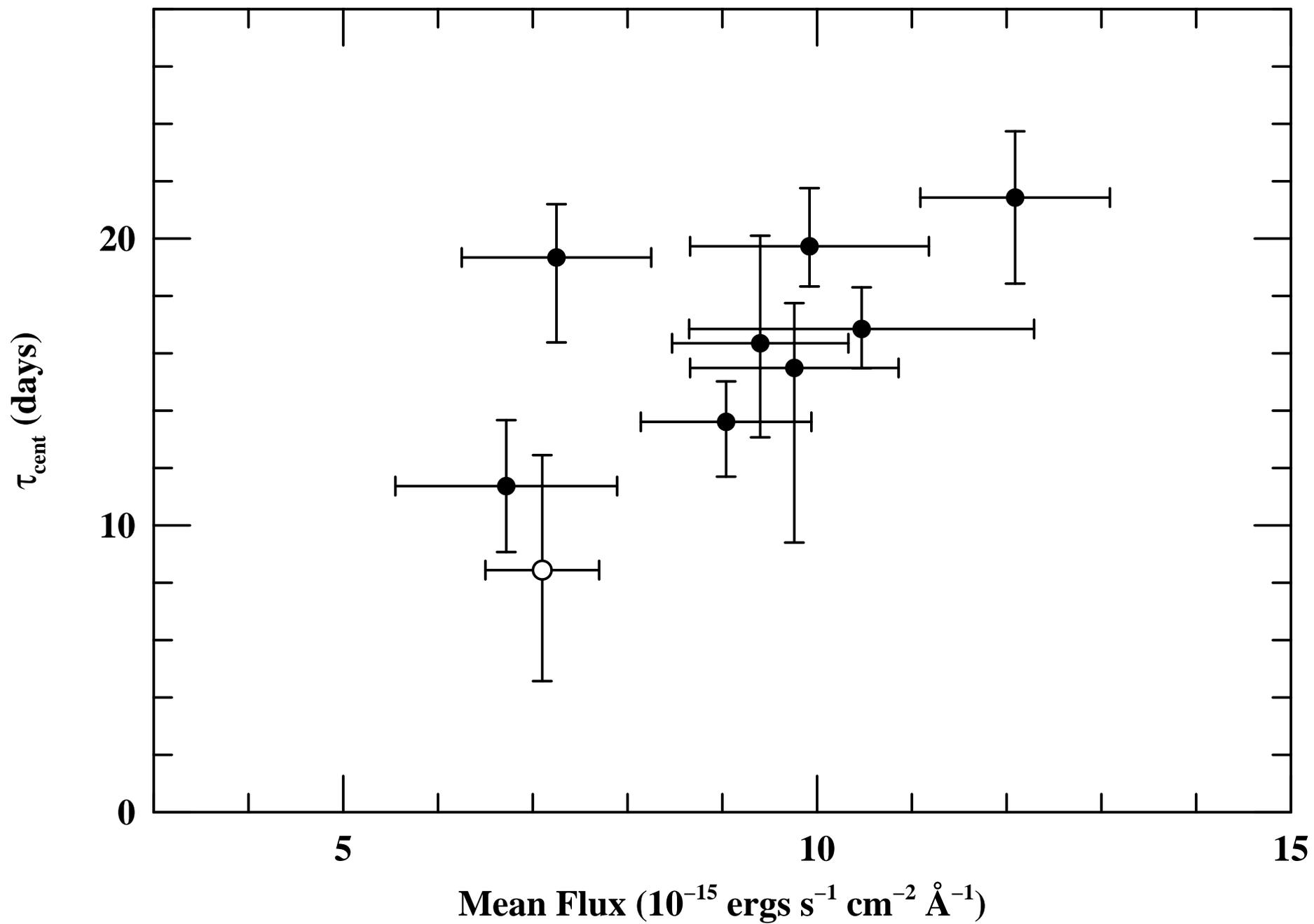
\end{document}